\documentstyle[epsfig]{mn}

\begin{document}

\title{On the cosmological evolution of the FRII radio source population}

\author[Christian R. Kaiser \& Paul Alexander]{Christian R. Kaiser$^{1,2}$\thanks{email: c.kaiser1@physics.oxford.ac.uk} and Paul Alexander$^1$\\
$^1$ MRAO, Cavendish Laboratory, Madingley Road, Cambridge, CB3 0HE,
UK\\ $^2$ University of Oxford, Department of Physics, Nuclear Physics
Laboratory, Keble Road, Oxford, OX1 3RH, UK }

\maketitle

\begin{abstract}
We present an analytical model for the cosmological evolution of the
FRII source population. Based on an earlier model for the intrinsic
radio luminosity - linear size evolution of these objects, we
construct theoretical source samples. The source distributions in the
radio power - linear size plane of these samples are then compared
with that of an observed flux-limited sample. We find that the source
parameters determining the radio luminosity of FRII objects can not be
independent of each other. The best-fitting models predict the jet
power to be correlated either with the life time of the source or with
the shape of the density distribution of the source environment. The
latter case is consistent with the observed tendency of the most
luminous radio sources at high redshift to be located in richer and
more extended environments than their low redshift counterparts. We
also find evidence for a class of FRII sources distinctly different
from the main population. These sources are extremely old and/or are
located in very underdense environments. The luminosity function of
FRII sources resulting from the model is in good agreement with
previous results for high luminosity sources. The apparent
luminosity evolution of the radio luminosity function is not reproduced
because of the high flux limit of the used comparison sample. The
cosmological evolution of the median linear size of FRII sources is
found to be mild.
\end{abstract}


\section{Introduction}

This is the third in a series of papers presenting an analytical model
for the evolution of FRII radio sources. In Kaiser \& Alexander
(1997)\nocite{ka96b}, hereafter KA, we developed a model for the
general dynamics of these sources. The intrinsic evolution of the
radio luminosity of a given radio source as a function of its linear
size was investigated in Kaiser, Dennett-Thorpe \& Alexander
(1997)\nocite{kda97a}, hereafter KDA. In this paper we use the model
for the intrinsic radio luminosity-linear size evolution of FRII
sources of KDA to constrain the cosmological evolution of these
sources and their progenitors.

The radio power (P) - linear size (D) diagram was introduced by
Shklovskii (1963)\nocite{is63} as a powerful tool to investigate the
evolution of extragalactic radio sources. Baldwin (1982)\nocite{jb82}
pointed out that the P-D diagram is analogous to the
Hertzsprung-Russel diagram for stars. However, there is a strong
correlation between the radio luminosity and the redshift of sources
in flux-limited samples. This Malmquist bias implies that the source
distribution in the P-D plane of such samples is a result of the
intrinsic evolution of individual sources and the cosmological
evolution of the source population as a whole. KDA point out that the
scarcity of radio sources with linear sizes greater than roughly 1 Mpc
can be partly explained by the steepening of the evolutionary tracks
through the P-D diagram of FRII sources caused by the energy losses of
the relativistic electrons in their cocoons due to inverse Compton
scattering of the Cosmic Microwave Background Radiation (CMBR). Since
the energy density of the CMBR increases with redshift, the cut-off in
linear size above which only very few sources are found will put
important constraints on the cosmological evolution of the radio
source population. 

It is not clear why AGN formation occurs and how jet activity in
galaxies is subsequently triggered. It is therefore also unknown
whether all massive galaxies in the universe go through a phase of jet
activity as part of their evolution or whether only those galaxies
involved in violent processes like galaxy mergers can become active.

For the purpose of the analysis presented in this paper we will simply
assume that there is a population of progenitors of radio sources
without specifying exactly what they are.  The only property these
progenitors must have, is that they become active with a certain
probability at some cosmological epoch, as described by the `birth
function', and then turn into a radio source whose intrinsic
luminosity evolution is determined by the properties of the jets and
the environment the progenitors are located in.  With this assumption
there are then two possibilities how cosmic evolution in the number
density of radio sources can occur.  Either all progenitors existing
at a given redshift become active at some point in their life time and
the evolution reflects the number of progenitors created at or before
this redshift; or the progenitors are created very early in the
evolution of the universe, have very long life times and jet activity
is triggered by a process not intrinsic to the progenitor. In this
case any evolution in the number density of radio sources indicates
that the probability for this process to occur is varying with
redshift. For convenience we will use only the latter interpretation
in the formulation of the models in the following sections. However,
it should be born in mind that there is no way inherent in the model
to decide which interpretation is the more likely.

\begin{figure*}
\centerline{\epsfig{file=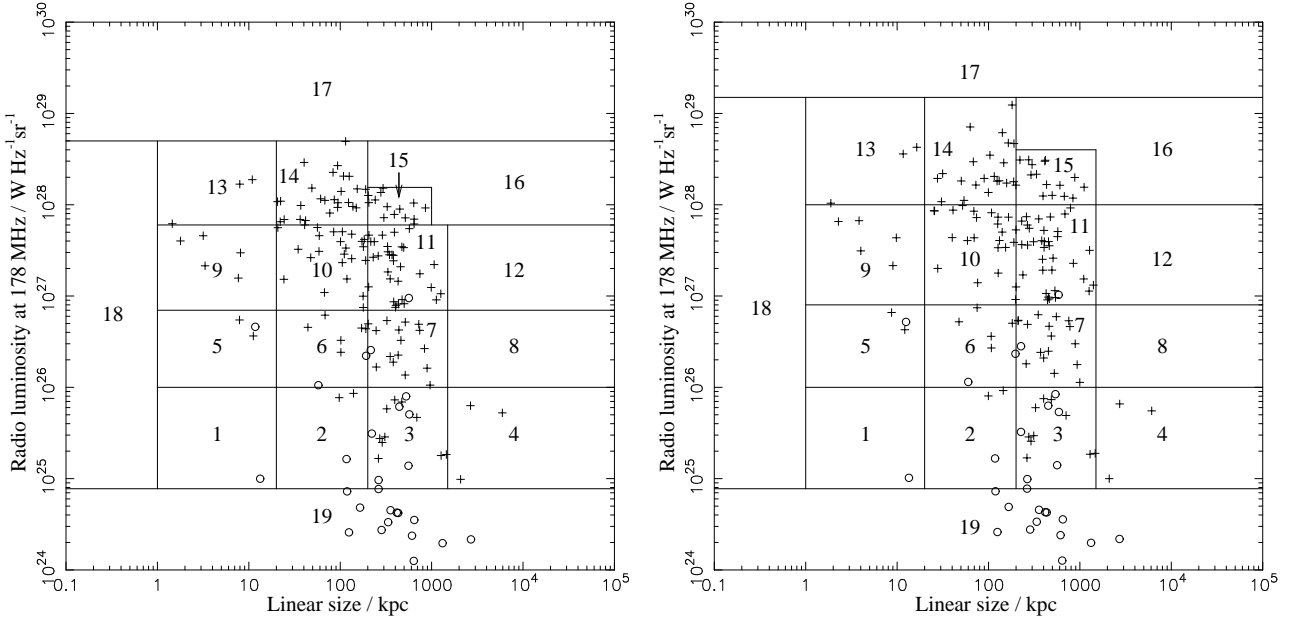, height=17cm, angle=270}}
\caption{\footnotesize The source distribution of the LRL sample in
the P-D plane. Left: $\Omega _o =1$, right: $\Omega _o=0$. Crosses:
FRII-type radio galaxies and quasars, circles: FRI-type radio
galaxies. The binning of the P-D plane used to compare model
predictions to the LRL sample is shown.}
\label{fig:pdbin}
\end{figure*}

Assuming a birth function and a set of distribution functions
characteristic of the environments of the progenitors, it is possible
with the aid of the model described by KDA to predict a P-D diagram
which accounts for observational selection effects. To constrain the
models the predicted P-D diagram is then compared directly with the
observed distribution of sources. Knowledge of the selection effects,
and hence the use of a complete observational sample is essential for
a proper statistical comparison. The only published sample of
extragalactic radio sources with the completeness required for this
analysis is the sample of Laing {\em et al.}  (1983)\nocite{lrl83},
hereafter LRL. In the original form the authors showed that the LRL
sample is 96\% complete to a flux limit of 10 Jy at 178 MHz for
sources with angular sizes less than $10'$. Riley (1989)\nocite{jr89}
showed that the sample is complete even for extended sources down to
the surface brightness limit of the 6C survey of 120 mJy per beam
(Baldwin {\em et al.}  1985\nocite{bbhjwww85}). The sample subtends a
solid angle of roughly 4.1 sr and contains 173 radio sources of which
30 are of type FRI. The remaining 143 sources are either quasars (43
sources) or radio galaxies (100 sources) with FRII morphology. We will
assume here that quasars and FRII radio galaxies form one class of
objects distinguished only by a different viewing angle at which they
are observed (e.g. Barthel 1989\nocite{pb89}). The model used here for
the calculation of the radio emission of FRII sources does not include
the emission of the radio core. The low selection frequency of LRL
ensures that relativistically beamed radio emission from the radio
core of FRII sources does not contribute significantly to the overall
flux. However, two sources in the sample, 3C 345 and 3C 454.3, may
have been included in the sample only because of their boosted core
emission. The changes to the statistical properties of the LRL sample
introduced by excluding these two sources are negligible. The source
distribution of the LRL sample in the P-D plane is shown in Figure
\ref{fig:pdbin}.

Once the most likely birth function and distribution functions of the
progenitor properties is found, it is straightforward to derive the
luminosity function of FRII sources from the model. Peacock
(1985)\nocite{jp85} and Dunlop \& Peacock (1990)\nocite{dp90} have
used a large data base of deep radio observations to determine the
luminosity function of all radio galaxies and quasars. Their free
modelling approach allows the shape of the luminosity function to be
constrained but does not explain why it has this shape. We show that
the model described here is in agreement with their results and allows
some conclusions as to the origin of the form of the luminosity
function.

The median linear size of luminous sources in the LRL sample appears
to be shorter than that of low luminosity sources (see Figure
\ref{fig:pdbin}). Because of the correlation between luminosity and
redshift in flux-limited samples this effect can also be interpreted
in terms of a decrease in median linear size, $D_{med}$, with redshift
(e.g. Masson 1980\nocite{cm80} and Macklin 1982\nocite{jm82}). If
interpreted in this way, this effect can be modelled with a power law,
$D_{med} \propto (1+z)^{n_D}$. Several groups have tried to determine
$n_D$ and find very different values. Oort {\em et al.}
(1987)\nocite{okw87}, Singal (1988)\nocite{as88}, Kapahi
(1989)\nocite{vk89} and Subrahmanian \& Swarup (1990)\nocite{ss90}
find $n_D \sim 3$ while Eales (1985)\nocite{se85} and Neeser {\em et
al.} (1995)\nocite{ner95} derive $n_D \sim 1.3$. Neeser {\em et al.}
(1995) showed that the strong evolution found in some samples could be
caused by a selection effect in which FRI sources are mistakingly
included in samples of FRII sources leading to an overestimate of the
median linear size at low redshift. However, the observation that the
maximum and the median linear sizes of high redshift and therefore
luminous sources are shorter than those of their lower redshift and
less luminous counterparts must be explained by any model predicting
the radio source distribution in the P-D plane.

In Section 2 we construct samples of radio sources predicted by
various source distribution functions and compare their distribution
in the P-D plane with that of the LRL sample. The luminosity function
of FRII sources predicted by the best-fitting models is derived in
Section 3. In Section 4 we investigate the cosmological evolution of
the median linear size of FRII objects as predicted by our model. A
discussion of the implications of the model for the cosmological
evolution of the radio source progenitor population is presented in
Section 5. Throughout this paper we are using Friedmann world models
with $\Lambda =0$ and $H_o =50$ km s$^{-1}$ Mpc$^{-1}$.

\section{Theoretical source distribution in the P-D plane}

\begin{figure*}
\centerline{\epsfig{file=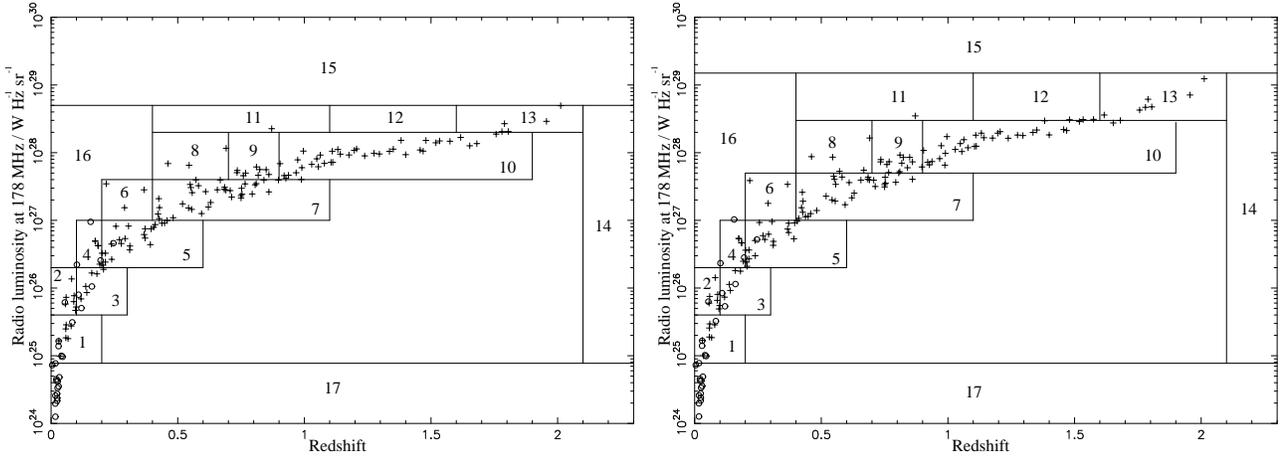, height=17cm, angle=270}}
\caption{\footnotesize The source distribution of the LRL sample in
the P-z plane. Left: $\Omega _o =1$, right: $\Omega _o =0$. Crosses:
FRII-type radio galaxies and quasars, circles: FRI-type radio
galaxies. The binning of the P-z plane used to compare model
predictions to the LRL sample is shown.}
\label{fig:pzbin}
\end{figure*}

In this section we describe the construction of theoretical P-D
diagrams and the method employed to compare them with the observed
distribution of FRII sources in the P-D plane.

\subsection{The source distribution function}

The model presented in KA and KDA for the radio luminosity and
linear size evolution of a given radio source depends on a variety of
source and environment parameters. If the density distribution of the
gaseous environment of the source, $\rho _x$, is approximated by a
King (1972)\nocite{ik72} profile,

\begin{equation}
\rho _x = \rho _o \left[ 1+ \left( \frac{r}{a_o} \right) ^2 \right]
^{-\frac{\beta}{2}},
\label{king}
\end{equation}

\noindent where $r$ is the radial distance from the centre of this
distribution, these defining parameters are: The jet power, $Q_o$, the
initial opening angle of the jet, $\theta$, the central density of the
density distribution of the source environment, $\rho _o$, the core
radius of this density distribution, $a_o$, and the exponent of the
power law describing this distribution, $\beta$. Since it is not
straightforward to measure $\theta$ directly in observations, we will
use the aspect ratio of the cocoon, $R$, i.e. the ratio of the length
of the cocoon and its width, which is related to $\theta$ by
$c_2/\theta ^2 \approx 4 R^2$ in the case of a ram-pressure confined
cocoon, where $c_2$ is a known constant (see KDA). 

We assume that both, the energy density of the particles in the cocoon
and that of the energy density of the magnetic field in the same
region have relativistic equations of state which corresponds to case
1 in KDA. It was shown in that paper that the specific choice of the
equation of state in both cases does not significantly influence the
shape of the resulting evolutionary tracks of a given source through
the P-D diagram. Note also, that our choice implies equipartition
between the energy density of the particles and that of the magnetic
field in the whole cocoon. We also assume that the power law
describing the energy spectrum of the relativistic electrons in the
cocoon of all model sources extends from a Lorentz factor $\gamma =1$
to $\gamma \rightarrow \infty$ and that the exponent of the power law
is $p=2.25$, implying mildly relativistic bulk velocities in the jets
with a Lorentz factor of roughly 2 (Heavens \& Drury
1988\nocite{hd88}).

For a given choice of source and environmental parameters, together
with the source age, $t_l$, the redshift of the source, $z_p$, and the
viewing angle between the jet axis and the line of sight, $\alpha _v$,
the model developed in KDA yields a point on the P-D plane. In order
to construct a complete theoretical P-D diagram one ideally needs to
know the distribution function of each of the parameters mentioned
above. Furthermore, the distribution functions of various parameters
may not be independent, e.g. the sources with more powerful jets are
also situated in higher density environments. For simplicity we assume
initially that the distribution functions required in the model are
independent. The number density of FRII sources in the universe as a
function of jet and environmental parameters and of source age, $t_l$,
can then be written as

\begin{eqnarray}
d^6 \rho & = & \rho _{tot} \ N_1 (\log _{10} Q_o) \, d(\log _{10} Q_o) \ N_2 (R) \, dR \nonumber\\
& \times & N_3 (\log _{10} \rho _o) \, d(\log _{10} \rho _o) \ N_4 (a_o) \, da_o \nonumber\\
& \times & N_5 (\beta) \, d\beta \ N_6 (t_l) \, dt_l,
\label{weight}
\end{eqnarray}

\noindent where $\rho _{tot}$ is the total number density of FRII
sources in the universe and $N_i$, where $i$ runs from 1 to 6, are the
distribution functions of the jet and environmental parameters. All
$N_i$ are normalised to yield 1 when integrated over the entire range
of the respective parameter.

In the chosen cosmology ($\Lambda =0$) the progenitors of radio
sources of age $t_l$ observed at redshift $z_p$ became active at a
cosmological epoch corresponding to

\begin{equation} 
z_s = \left[ \left( z_p +1 \right) ^{-\frac{3}{2}} - \frac{3 \,
H_o}{2} \, t_l \right] ^{-\frac{2}{3}} \, - 1,
\end{equation}

\noindent for $\Omega _o = 1$ and

\begin{equation}
z_s = \frac{z_p +1}{1-H_o \, t_l \, \left( z_p +1 \right)} \, -1,
\end{equation}

\noindent for $\Omega _o = 0$. Let the probability for a progenitor to
`switch on' as a function of redshift, the `birth function' of FRII
sources, be $N_7 (z_s) dz_s$. The distribution function of the ages of
radio sources, $t_l$, observed at redshift $z_p$ then is

\begin{eqnarray}
N_6 (t_l) \, dt_l & = & N_7 (z_s) \, \frac{dz_s}{dt_l} \, dt_l \nonumber\\
& = & N_7' (t_l,z_p) \, \frac{H_o}{\left[ \left( z_p +1 \right) ^{-\frac{3}{2}} - \frac{3 H_o}{2} \, t_l \right] ^{\frac{5}{3}}} \, dt_l,
\label{lt1}
\end{eqnarray}

\noindent for $\Omega _o =1$ and

\begin{equation}
N_6 (t_l) \, dt_l = N_7' (t_l, z_p) \, H_o \, \left[ \frac{ z_p +1}{1- H_o \, t_l \, \left( z_p +1 \right)} \right] ^2 \, dt_l,
\label{lt0}
\end{equation}

\noindent for $\Omega _o =0$. If not stated otherwise we assume the
remaining distribution functions $N_i(A)$, with $i=1$\ldots 5, to be
uniform;

\begin{equation}
N_i (A) \, dA = \frac{f(A_{max}, A_{min})}{A_{max} -A_{min}} \, dA,
\label{normtophat}
\end{equation}

\noindent where

\begin{equation} f(A_{max},A_{min}) = \left\{ 
\begin{array}{ll} 1 &
\mbox{; $A_{min} \le A \le A_{max}$} \\
0 & \mbox{; $A<A_{min}$ or $A_{max} < A$}
\end{array} \right. .
\label{tophat}
\end{equation}

$N_6(t_l) dt_l$ in equations (\ref{lt1}) and (\ref{lt0}) is the
fraction of all FRII sources in the universe with an age $t_l$ at a
cosmological epoch corresponding to redshift $z_p$. Of these only the
sources located at coordinate distances between $R_o r (z_p)$ and $R_o
r (z_p +dzp)$ are observable. Assuming that the survey which we will
use to compare our model predictions with subtends a solid angle of
$\Omega$ and that the highest redshift at which progenitors become
active is $z_{max}$, this fraction is found to be

\begin{eqnarray}
\frac{dV(z_p)}{V(z_{max})} & = & \frac{\Omega \left[ R_o r (z_p) \right]
^2 \, d(R_o r)}{\Omega / 3 \left[ R_o r (z_{max}) \right] ^3} \nonumber\\
& = & \frac{3}{2} \, \sqrt{ \frac{ \left( z_{max} +1 \right) ^3}{ \left( z_p+1 \right) ^5}} \, \frac{\left( \sqrt{z_p +1} -1 \right) ^2}{\left( \sqrt{z_{max} +1} -1 \right) ^3} \, dz_p,
\label{vol1}
\end{eqnarray}

\noindent in the case of $\Omega _o =1$ and

\begin{eqnarray}
\frac{dV(z_p)}{V(z_{max})} & = & \frac{3 \, \left( z_{max} +1 \right) ^3}{z_{max}^3 \, \left( 1 + \frac{1}{2} z_{max} \right) ^3} \nonumber\\
& \times & \frac{z_p^2 \, \left( 1+\frac{1}{2} z_p \right) ^2 \, \left( 1 +z_p + \frac{1}{2} z_p^2 \right)}{\left( 1+ z_p \right) ^4 } \, dz_p,
\label{vol0}
\end{eqnarray}

\noindent for $\Omega _o =0$. 

From the jet and environmental properties it is possible to determine
the linear size of a given radio source for any source age, $t_l$, and
the observed linear size is determined by projecting onto the plane of
the sky using a distribution function $\sin \alpha _v \, d\alpha _v$
for viewing angles between $\alpha _v$ and $\alpha _v + d\alpha _v$.

With these results the relative number of sources within the
population of given jet and environmental parameters observed at
redshift $z_p$ and at a viewing angle $\alpha _v$ are
determined. Computationally, this is achieved by dividing the range
over which a property $A$ varies into small intervals $\Delta A$,
which are then identified with the differentials $dA$ in equation
(\ref{weight}). The model for the radio emission of the cocoon
described in KDA us then used to convert the chosen source properties
into a linear size and a radio luminosity at the frequency in the
rest-frame of the source, $P_{\nu (z_p+1)}$, corresponding to an
observing frequency, $\nu$. The flux density which would be measured
from the source is then

\begin{equation}
S_{\nu}=\frac{P _{\nu (z_p+1)}}{(R_o r)^2 \, (1+z_p)},
\label{flux}
\end{equation}

\noindent where $R_o r$ is the coordinate distance of the source. If
this flux density is greater then the flux density limit of the
observational sample the source is included in the model sample with
the appropriate weight given by equation (\ref{weight}) and multiplied
by $dV(z_p)/V(z_{max}) \, \sin \alpha _v \, d\alpha _v$.

Objects in flux-limited samples of observed radio sources span a large
range in redshift. To obtain radio luminosities for these objects at
the same frequency in the rest-frame of the individual sources it is
usually assumed that the spectrum of radio galaxies is well
represented by a power law, $P'_{\nu} \propto \nu ^{-\alpha}$, where
the spectral index, $\alpha$, is determined by measuring the flux of
the radio galaxy at a second frequency. For the LRL sample this second
frequency is 750 MHz. With this assumption equation (\ref{flux}) can
be inverted to give

\begin{equation}
P'_{\nu} = S_{\nu} \, (R_o r)^2 \, (1+z_p)^{1+\alpha}.
\label{flux2}
\end{equation}

\noindent In order to mimic this procedure we calculate the radio
luminosity of every source in the model samples at 178 MHz and 750 MHz
from the model of KDA and derive the `observed' spectral index,
$\alpha$. With this and using equation (\ref{flux2}) we then calculate
the `observed' radio luminosity, $P'(\nu)$, which is then used to
determine the location of the source on the P-D plane. Note, that
$P'_{\nu}$ is independent of the value of $\Omega _o$ while the flux
density calculated from equation (\ref{flux}) depends on the chosen
cosmology. The sets of sources within the model samples are therefore
different for different cosmologies.

Unless most of the radio emission from large sources is concentrated
in the hot spots, it can be distributed over a large projected area on
the sky. Even if the total flux of such sources is above the flux
limit of a given sample, the sources may not be included in the sample
because their surface brightness is below the detection limit of the
telescope used to compile the survey from which the sample is
drawn. In the model presented here we assume that the surface
brightness of a given source is equal to its total luminosity divided
by the projected area of sky covered by the cocoons. If the calculated
surface brightness is below the surface brightness limit of the 6C
sample, which was used to check the completeness of the LRL sample
(Riley 1989\nocite{jr89}), the source is not included in the model
sample. Using this selection criterion we may omit some sources from
our model samples which should be included because their radio
emission is less smoothly distributed and their hot spots may be
bright enough to be detected. However, in all the models presented
here, the `number' of sources excluded because of insufficient surface
brightness was of order $10^{-4}$ at most, which is negligible when
compared to the number of sources, $10^{-2}$ or higher, in the
relevant bin which were included in the model sample.

\subsection{Model parameters\label{sec:mod}}

The range over which the source parameters used in the model vary can
in principle be deduced from observations, however, the constraints
obtained in this way are incomplete and we will show later that there
may also be considerable variation with redshift.

For the jet powers of FRII sources with redshifts below $z=1$ Rawlings
\& Saunders (1991)\nocite{rs91} find $10^{37}$ W $\le Q^{rs} \le
10^{40}$ W from minimum energy arguments and $\Omega _o =0$. For
$\Omega _o=1$ these limiting values of $Q^{rs}$ are somewhat lower but
even for $z=1$ the correction factor necessary is less than 2. KDA
point out that another correction due to an underestimate of the
expansion work done by the cocoon by Rawlings \& Saunders
(1991)\nocite{rs91} and to the potential presence of thermal material
in the cocoon is necessary. Their equation (18) assumes that the rate
at which energy is lost by the cocoon as expansion work as predicted
by the model of KA is equal to half the jet power found by Rawlings \&
Saunders (1991)\nocite{rs91}. Because of the method used by Rawlings
\& Saunders (1991) it is more accurate to assume the rate at which
energy is stored in the cocoon as predicted by KA to be equal to
$Q^{rs}/2$. This implies that $Q^{rs}$ is an overestimate of the jet
power. When the additional energy of the thermal particles required by
the model in KDA is also taken into account, we find that the
correction to $Q^{rs}$ is of order unity. We will therefore initially
use $Q_{o,min} =10^{37}$ W and $Q_{o,max} = 10^{40}$ W. A further
discussion of this point is deferred to Section 3.

The source distribution of the LRL sample in the P-z plane (see Figure
\ref{fig:pzbin}) implies that at every redshift there are more FRII
sources of low radio luminosity in the universe than very luminous
objects. We therefore introduce an exponential distribution of jet
powers, $Q_o$;

\begin{eqnarray}
\lefteqn{N_1(\log _{10} Q_o) \, d\left(\log_{10} Q_o
\right)=}\nonumber\\ & & \frac{ \lambda _q \, e^{-\lambda _q \left(
\log _{10} Q_o - \log_{10} Q_{o,min} \right)} }{1-e^{-\lambda _q
\left( \log _{10} Q_{o,max} -\log _{10} Q_{o,min} \right)}} \, d\left(
\log _{10} Q_o \right),
\label{powdis}
\end{eqnarray}

\noindent where $\lambda _q$ varies from 0 to 10 in steps of 1.0. For
$\lambda _q= 0$ we use a uniform distribution of $\log _{10} Q_o$ as
described by equation (\ref{normtophat}).

For the aspect ratio of the cocoon, $R$, we use a uniform distribution
with $1.3 \le R \le 6.0$, taken from Leahy \& Williams
(1984)\nocite{lw84}. The environments of FRII sources are either the
IGM in clusters of galaxies or the atmospheres of the host galaxy
itself in the case of small linear sizes or isolated
objects. Canizares {\em et al.} (1987)\nocite{cft87} find $4\cdot
10^{-23}$ kg/m$^3$ $\le \rho _o \le 5\cdot 10^{-21}$ kg/m$^3$ for the
central density of individual galaxies. The core radius of the density
profile in these objects is $0.01$ kpc $\le a_o \le 2$ kpc with a
median value $\bar{a}_o \approx 1.0$ kpc. For clusters of galaxies
Jones \& Forman (1984)\nocite{jf84} find $7\cdot 10^{-25}$ kg/m$^3$
$\le \rho _o \le 5\cdot 10^{-23}$ kg/m$^3$ and $30$ kpc $\le a_o \le
1100$ kpc with $\bar{a} _o \approx 260$ kpc. For simplicity we will
calculate theoretical P-D diagrams for individual sources and sources
in clusters separately and we will also assume that the core radius,
$a_o$, for all model sources is equal to the respective median value
$\bar{a} _o$. This implies $N_4(a_o) \, da_o = \delta (a_o=\bar{a}_o)
\, da_o$ in equation (\ref{weight}). For the central density we use a
uniform distribution with $\rho _{o,min} = 5 \cdot 10^{-23}$ kg/m$^3$
and $\rho _{o,max} = 5\cdot 10^{-21}$ kg/m$^3$ for isolated sources
and $\rho _{o,min} = 5 \cdot 10^{-25}$ kg/m$^3$ and $\rho _{o,max} =
5\cdot 10^{-23}$ kg/m$^3$ for sources in galaxy clusters.

In some of the following models we allow for cosmological evolution of
the central density, $\rho_o$. If the properties of the environments
of the progenitors only depend on redshift but not on any other source
property, i.e. the jet power, they should all evolve in very similar
ways. The central density, $\rho _o$, will in this case be
proportional to the density of the gas filling the universe at the
time when the progenitor and its environment decoupled from the Hubble
flow. In this simplified picture we therefore expect that $\rho _o
\propto (1+z)^{n_d}$, with $n_d \le 3$.

The model for the radio emission of the cocoons of FRII sources
described in KDA requires an external density distribution with $\rho
_x \propto r^{-\beta}$. For distances from the centre of the density
distribution, $r$, greater than a few core radii, $a_o$, the
approximation $\rho _x = \rho _o (r/a_o)^{-\beta}$ provides a good fit
to equation (\ref{king}). However, within the core radius this is not
the case and sources in the centre of galaxy clusters will spend a
considerable part of their life time in this region. We therefore
split the external density profile into three different regimes: $\rho
_x = \rho _o$ for $0 \le r \le a_o/2$, $\rho _x = \rho _o/ \, (2 \,
r/a_o)^{-\beta /2}$ for $a_o/2 \le r \le 2 a_o$ and $\rho _x = \rho _o
\, (r/a_o)^{-\beta}$ for $r > 2 a_o$. During its life time a given
radio source is changing from one density regime into the next when
the linear sizes of its cocoons, $L_j$, reach $a_o/2$ and $2 a_o$
respectively. Let the cocoons of a given radio source reach a linear
size of $a_o/2$ each at an age of $t_l
(L_j=a_o/2)=t_{l,0}(L_j=a_o/2)$. During this time the external density
profile is assumed to be $\rho _x = \rho _o$. If the source was
located in a density regime with $\rho _x = \rho _o \, (2 \,
r/a_o)^{-\beta /2}$ while expanding to this linear size, its age would
be $t_{l,\beta /2}(L_j=a_o/2)\ne t_{l,0}(L_j=a_o/2)$. To calculate the
correct radio luminosity using the model presented in KDA once the
cocoons of the source become larger than $a_o/2$ and enter the second
density regime, we have to use a ficticious source age $t_{l,\beta/2}
(L_j > a_o/2) = t_l (L_j > a_o/2) - t_{l,0}(L_j=a_o/2)+t_{l,\beta
/2}(L_j=a_o/2)$ instead of the real source age $t_l (L_j > a_o/2)$. In
other words, once the linear size of the cocoons of the radio source
become larger than $a_o/2$, we treat the source for the calculation of
its radio luminosity {\em as if}\/ it had been located in an
environment with $\rho _x = \rho _o \, (2 \, r/a_o)^{-\beta /2}$ for a
time equal to the fictious source age. Sources with cocoons larger
than $2a_o$ entering the third density regime are treated in an
analogous way. The aspect ratio of the cocoon, $R$, is assumed to
remain constant when the source changes from one regime to the
next. We assume a uniform distribution of $\beta$ between 1.0 and 2.0.

The viewing angle, $\alpha _v$, is distributed according to $\sin
\alpha _v \, d\alpha _v$ over the range $0$ to $\pi /2$ radians and for
the source age we assume a maximum value of $t_{l,max} = 10^9$ years
which is the upper limit of observed spectral ages (Alexander \& Leahy
1987)\nocite{al87}. For sources reaching the maximum age we assume
that the jet ceases to supply the cocoon with energy. The subsequent
drop in radio luminosity occurs fast because of the adiabatic
expansion of the cocoons. Therefore we assume that the radio
luminosity of a source of age $t_{l,max}$ drops to zero
instantaneously.

The ratio of the number of progenitors becoming FRII sources with
powerful jets to that of objects with weaker jets is given by equation
(\ref{powdis}) and is constant for all redshifts. The distribution of
observed sources within the LRL sample in the P-z plane (see Figure
\ref{fig:pzbin}) therefore suggests that the total number density of
FRII sources in the universe is higher at higher redshift. To allow
for this effect we introduce a birth function of the form $N_7 (z_s)
\, dz_s \propto (z_s+1)^n \, dz_s$. The distribution function of the
age of sources observed at redshift $z_p$ then becomes

\begin{eqnarray}
N_7'(t_l,z_p) & = & \frac{n+1}{\left( z_{max} +1 \right) ^{n+1} -1}
 \nonumber\\ 
& \times & \left[ \left( 1+z_p \right) ^{-\frac{3}{2}} - \frac{3\,
 H_o}{2} \, t_l \right] ^{-\frac{2}{3}\, n},
\label{birth1}
\end{eqnarray}

\noindent for $\Omega =1$ and 

\begin{eqnarray}
N_7'(t_l,z_p) & = & \frac{n+1}{\left( z_{max} +1 \right) ^{n+1} -1}
 \nonumber\\ 
& \times & \left\{ \frac{z_p +1}{\left[ 1- H_o \, t_l \,
 \left( z_p +1 \right) \right]} \right\} ^n,
\label{birth0}
\end{eqnarray}

\noindent for $\Omega =0$. In both cases we investigate values of $n$
between 0 and 10. Since there are no sources in the LRL sample beyond
a redshift of roughly 2.1 we will use $z_{max} =5$ as the cut-off for
the birth function. This ensures that even for $\Omega =0$ the first
generation of radio sources which became active at a cosmological
epoch corresponding to $z_s =5$ has reached the maximum source age at
the cosmological epoch corresponding to $z_p =2.1$.

\subsection{$\chi ^2$-analysis}

The comparison of the theoretical predictions for the distribution of
radio sources in the P-D diagram with the LRL sample is done by
splitting the P-D plane into 19 bins as shown in Figure
\ref{fig:pdbin}. The number of FRII-type sources for the LRL sample in
each of the bins is given in the first column of Tables
\ref{tab:sourcenumber1} and \ref{tab:sourcenumber0}. Note, that the
FRI-type objects in the LRL sample are not included in the binning
process.

\begin{table*}
{\centering
\begin{tabular}{ccrrrrrrrrrrrrrr}
 & & \multicolumn{2}{c}{A} & \multicolumn{2}{c}{B} &
\multicolumn{2}{c}{C} & \multicolumn{2}{c}{D} & \multicolumn{2}{c}{E}
& \multicolumn{2}{c}{F} & \multicolumn{2}{c}{G} \\[-1ex]
& & \multicolumn{2}{c}{\rule{15ex}{0.1ex}} & \multicolumn{2}{c}{\rule{15ex}{0.1ex}} & \multicolumn{2}{c}{\rule{15ex}{0.1ex}} & \multicolumn{2}{c}{\rule{15ex}{0.1ex}} & \multicolumn{2}{c}{\rule{15ex}{0.1ex}} & \multicolumn{2}{c}{\rule{15ex}{0.1ex}} & \multicolumn{2}{c}{\rule{15ex}{0.1ex}}\\
bin & LRL & {\em
gal.} & {\em cl.} & {\em gal.} & {\em cl.} & {\em gal.}
& {\em cl.} & {\em gal.} & {\em cl.} & {\em gal.} & {\em
cl.} & {\em gal.} & {\em cl.} & {\em gal.} & {\em
cl.}\\ 
\hline
1 & 0 & 0.73 & 0.27 & 0.66 & 0.16 & 0.54 & 0.19 & 0.61 & 0.21 & 0.61 & 0.21 & 0.34 & 0.13 & 0.35 & 0.15\\
2 & 2 & 6.78 & 6.75 & 5.95 & 3.31 & 4.81 & 4.05 & 6.06 & 4.67 & 5.05 & 4.20 & 3.26 & 2.77 & 3.51 & 3.45\\
3 & 10 & 16.46 & 16.72 & 13.24 & 7.22 & 10.59 & 8.64 & 11.22 & 9.86 & 8.96 & 8.14 & 7.23 & 6.52 & 8.09 & 8.09\\
4 & 3 & 2.41 & 0.32 & 1.44 & 0.17 & 0.73 & 0.11 & 0.33 & 0.08 & 0.60 & 0.09 & 0.46 & 0.03 & 0.35 & 0.04\\
\rule{0ex}{3ex}5 & 2 & 0.67 & 0.19 & 1.11 & 0.41 & 1.00 & 0.46 & 1.28 & 0.44 & 1.24 & 0.58 & 0.90 & 0.47 & 1.35 & 0.62\\
6 & 6 & 5.06 & 6.35 & 6.72 & 7.85 & 5.55 & 8.74 & 8.24 & 8.20 & 7.23 & 9.79 & 7.32 & 8.91 & 9.46 & 10.30\\
7 & 16 & 12.64 & 19.06 & 12.26 & 15.95 & 9.17 & 16.46 & 11.49 & 14.29 & 11.04 & 15.61 & 12.65 & 14.89 & 13.81 & 16.32\\
8 & 0 & 4.38 & 1.21 & 1.98 & 0.49 & 0.99 & 0.28 & 0.45 & 0.13 & 1.0 & 0.18 & 0.32 & 0.00 & 0.28 & 0.00\\
\rule{0ex}{3ex}9 & 5 & 1.11 & 0.32 & 1.60 & 0.91 & 2.65 & 1.38 & 2.89 & 1.34 & 4.11 & 2.35 & 3.38 & 2.07 & 3.76 & 2.12\\
10 & 23 & 9.55 & 9.22 & 11.25 & 14.52 & 18.47 & 21.71 & 22.37 & 20.50 & 25.64 & 27.25 & 28.03 & 31.20 & 26.05 & 26.60\\
11 & 32 & 23.95 & 29.86 & 22.02 & 28.89 & 34.00 & 39.41 & 33.44 & 35.57 & 35.87 & 32.39 & 34.80 & 34.43 & 32.08 & 26.19\\
12 & 0 & 11.03 & 5.03 & 3.28 & 0.92 & 2.10 & 0.49 & 0.73 & 0.25 & 1.33 & 0.19 & 0.12 & 0.00 & 0.25 & 0.00\\
\rule{0ex}{3ex}13 & 3 & 2.74 & 0.54 & 2.88 & 2.04 & 5.04 & 1.82 & 4.24 & 2.14 & 5.87 & 3.88 & 4.16 & 2.07 & 4.52 & 3.35\\
14 & 27 & 19.45 & 12.46 & 19.79 & 23.75 & 21.51 & 18.54 & 21.48 & 22.01 & 20.86 & 25.23 & 25.94 & 24.46 & 22.47 & 28.78\\
15 & 14 & 12.39 & 14.57 & 11.79 & 14.09 & 24.83 & 20.64 & 18.30 & 23.30 & 12.80 & 12.91 & 13.55 & 13.65 & 16.32 & 17.03\\
16 & 0 & 13.71 & 20.24 & 22.57 & 17.74 & 0.76 & 0.09 & 0.17 & 0.06 & 0.05 & 0.02 & 0.53 & 1.37 & 0.12 & 0.00\\
\rule{0ex}{3ex}17 & 0 & 0.00 & 0.00 & 4.58 & 4.84 & 0.00 & 0.00 & 0.00 & 0.00 & 0.00 & 0.00 & 0.00 & 0.00 & 0.00 & 0.00\\
18 & 0 & 0.01 & 0.00 & 0.12 & 0.01 & 0.25 & 0.01 & 0.18 & 0.01 & 0.20 & 0.02 & 0.01 & 0.00 & 0.25 & 0.03\\
19 & 0 & 0.00 & 0.00 & 0.00 & 0.00 & 0.00 & 0.00 & 0.00 & 0.00 & 0.00 & 0.00 & 0.00 & 0.00 & 0.00 & 0.00\\
\hline
$\chi ^2$ & \ldots & 78.1 & 190.8 & 67.3 & 105.6 & 29.7 & 100.2 & 34.3 & 135.0 & 21.5 & 108.1 & 21.5 & 299.4 & 27.0 & 238.1\\
$\chi _4^2$ & \ldots & 78.3 & 171.3 & 66.3 & 58.2 & 23.0 & 26.5 & 12.8 & 24.9 & 12.0 & 11.4 & 7.7 & 17.1 & 6.9 & 12.7\\
\hline
\end{tabular}}
\caption{The number of sources in each P-D bin as defined in Figure
\ref{fig:pdbin} for the LRL sample and models A to G for $\bf \Omega
_o =1$. {\em gal.} and {\em cl.} indicate the use of the isolated
galaxy and the cluster density profile respectively. $\chi ^2$ is the
result of the $\chi ^2$-test for the respective model and $\chi _4^2$
is the result of the $\chi ^2$-test when omitting bin 4.}
\label{tab:sourcenumber1}
\end{table*}

\begin{table*}
{\centering
\begin{tabular}{ccrrrrrrrrrrrrrr}
 & & \multicolumn{2}{c}{A} & \multicolumn{2}{c}{B} &
\multicolumn{2}{c}{C} & \multicolumn{2}{c}{D} & \multicolumn{2}{c}{E}
& \multicolumn{2}{c}{F} & \multicolumn{2}{c}{G} \\[-1ex]
& & \multicolumn{2}{c}{\rule{15ex}{0.1ex}} & \multicolumn{2}{c}{\rule{15ex}{0.1ex}} & \multicolumn{2}{c}{\rule{15ex}{0.1ex}} & \multicolumn{2}{c}{\rule{15ex}{0.1ex}} & \multicolumn{2}{c}{\rule{15ex}{0.1ex}} & \multicolumn{2}{c}{\rule{15ex}{0.1ex}} & \multicolumn{2}{c}{\rule{15ex}{0.1ex}}\\
bin & LRL & {\em
gal.} & {\em cl.} & {\em gal.} & {\em cl.} & {\em gal.}
& {\em cl.} & {\em gal.} & {\em cl.} & {\em gal.} & {\em
cl.} & {\em gal.} & {\em cl.} & {\em gal.} & {\em
cl.}\\ 
\hline
1 & 0 & 0.70 & 0.15 & 0.69 & 0.17 & 0.60 & 0.19 & 0.53 & 0.21 & 0.61 & 0.18 & 0.39 & 0.21 & 0.37 & 0.16\\
2 & 2 & 6.23 & 3.88 & 5.98 & 3.47 & 5.17 & 3.89 & 5.35 & 4.55 & 5.13 & 3.42 & 3.73 & 4.78 & 3.72 & 3.27\\
3 & 10 & 13.01 & 9.60 & 11.25 & 7.77 & 9.77 & 8.59 & 10.17 & 10.05 & 9.16 & 6.88 & 8.41 & 11.08 & 8.34 & 7.62\\
4 & 3 & 1.14 & 0.18 & 0.73 & 0.18 & 0.64 & 0.11 & 0.27 & 0.08 & 0.54 & 0.13 & 0.54 & 0.04 & 0.39 & 0.05\\
\rule{0ex}{3ex}5 & 2 & 0.77 & 0.15 & 1.36 & 0.42 & 1.33 & 0.40 & 1.26 & 0.39 & 1.38 & 0.56 & 1.03 & 0.55 & 1.62 & 0.87\\
6 & 5 & 6.09 & 5.05 & 7.86 & 8.30 & 7.75 & 7.84 & 7.65 & 7.54 & 7.81 & 10.02 & 8.55 & 10.90 & 10.37 & 13.01\\
7 & 17 & 13.84 & 15.09 & 13.55 & 18.27 & 13.51 & 16.24 & 12.07 & 14.93 & 12.41 & 18.24 & 14.40 & 17.39 & 15.37 & 19.96\\
8 & 0 & 3.25 & 0.98 & 1.35 & 0.62 & 1.34 & 0.31 & 0.52 & 0.17 & 1.04 & 0.37 & 0.36 & 0.00 & 0.31 & 0.01\\
\rule{0ex}{3ex}9 & 5 & 1.50 & 0.37 & 2.36 & 1.02 & 3.06 & 1.22 & 3.03 & 1.33 & 3.88 & 2.42 & 3.17 & 2.00 & 4.23 & 2.87\\
10 & 23 & 13.92 & 11.25 & 15.88 & 17.15 & 20.78 & 19.54 & 23.03 & 21.14 & 24.11 & 28.52 & 28.08 & 29.15 & 26.72 & 30.97\\
11 & 36 & 33.08 & 37.13 & 28.69 & 36.69 & 37.56 & 38.94 & 38.97 & 40.38 & 34.38 & 35.64 & 34.77 & 28.30 & 30.14 & 27.69\\
12 & 0 & 11.84 & 7.30 & 2.97 & 1.45 & 3.10 & 0.82 & 1.38 & 0.46 & 1.74 & 0.45 & 0.16 & 0.00 & 0.27 & 0.00\\
\rule{0ex}{3ex}13 & 3 & 1.88 & 0.51 & 3.70 & 1.54 & 3.49 & 1.70 & 2.61 & 1.41 & 5.25 & 3.16 & 3.08 & 2.30 & 4.79 & 3.22\\
14 & 22 & 14.55 & 13.01 & 19.87 & 18.34 & 16.50 & 18.69 & 16.58 & 17.07 & 20.98 & 21.51 & 22.58 & 23.95 & 22.48 & 23.50\\
15 & 15 & 19.08 & 35.33 & 22.97 & 23.97 & 18.18 & 24.55 & 19.49 & 23.34 & 14.36 & 11.49 & 13.76 & 12.35 & 13.65 & 9.81\\
16 & 0 & 2.10 & 3.07 & 3.72 & 3.75 & 0.05 & 0.02 & 0.03 & 0.01 & 0.01 & 0.01 & 0.00 & 0.00 & 0.00 & 0.00\\
\rule{0ex}{3ex}17 & 0 & 0.00 & 0.00 & 0.00 & 0.00 & 0.00 & 0.00 & 0.00 & 0.00 & 0.00 & 0.00 & 0.00 & 0.00 & 0.00 & 0.00\\
18 & 0 & 0.01 & 0.00 & 0.15 & 0.00 & 0.16 & 0.01 & 0.12 & 0.00 & 0.20 & 0.01 & 0.01 & 0.00 & 0.29 & 0.03\\
19 & 0 & 0.00 & 0.00 & 0.00 & 0.00 & 0.00 & 0.00 & 0.00 & 0.00 & 0.00 & 0.00 & 0.00 & 0.00 & 0.00 & 0.00\\
\hline
$\chi ^2$ & \ldots & 47.1 & 180.4 & 32.1 & 80.9 & 22.1 & 100.4 & 39.5 & 139.6 & 21.2 & 75.9 & 18.3 & 227.0 & 25.6 & 205.9\\
$\chi _4^2$ & \ldots & 44.6 & 138.3 & 25.4 & 38.3 & 13.6 & 28.5 & 12.6 & 27.5 & 10.2 & 14.4 & 7.1 & 18.1 & 8.1 & 17.6\\
\hline
\end{tabular}}
\caption{The number of sources in each P-D bin as defined in Figure
\ref{fig:pdbin} for the LRL sample and models A to G for $\bf \Omega
_o =0$. {\em gal.} and {\em cl.} indicate the use of the isolated
galaxy and the cluster density profile respectively. $\chi ^2$ is the
result of the $\chi ^2$-test for the respective model and $\chi _4^2$
is the result of the $\chi ^2$-test when omitting bin 4.}
\label{tab:sourcenumber0}
\end{table*}

The way in which the P-D plane is binned and the number of bins will
influence how well particular model predictions agree with the
observed distribution for the LRL sample. Too many bins result in few
sources in each occupied bin and it is then difficult within the $\chi
^2$ analysis to distinguish between occupied bins and those which are
empty. As we have pointed out earlier, the maximum linear size of
radio sources for each radio luminosity and the maximum radio
luminosity within the whole sample are the strongest constraints on
the cosmological evolution of the progenitor population. We have
therefore chosen the bins in Figure \ref{fig:pdbin} to delineate the
source distribution of the LRL sample for large linear sizes and high
radio luminosities without increasing significantly the number of
bins.

For each set of model parameters for the progenitor population two
`samples' are calculated: one has external density profiles
characteristic of isolated galaxies and the other density profiles
appropriate to cluster environments. In all cases the flux density and
surface brightness limits match those of the LRL sample. These
`samples' are then normalised to contain in total the same number of
FRII--type objects as the LRL sample and binned in the way shown in
Figure \ref{fig:pdbin}. The normalisation determines the total number
density of FRII sources in the universe, $\rho _{tot}$, in equation
(\ref{weight}).

The distribution of sources over the P-D diagram resulting from the
model calculation is then compared with the one of the LRL sample
using a $\chi ^2$ test. The probability that the source distribution
of the LRL sample is a representation of a universe with the chosen
source distribution function, equation (\ref{weight}), increases with
decreasing $\chi ^2$.

Since the redshifts of all objects in the LRL sample are known it is
also possible to compare the model predictions with the distribution
of observed sources in the P-z plane. Figure \ref{fig:pzbin} shows the
objects in LRL in this plane and the 17 bins used for the $\chi
^2$-test here. The binning of the P-z plane is complicated by the
Malmquist bias. Most sources in the LRL sample and the model samples
are close to the line defined by the flux limit and it is difficult to
find bins in this plane that will quantify the scatter about this line
properly. Ideally one would like to use the two dimensional
Kolmogorov-Smirnov test to compare the models with the observed data
in the P-z plane. However, Peacock (1983)\nocite{jp83} showed that the
two dimensional KS test is unreliable if the two independent
variables, in this case P and z, are strongly correlated. Since the
main interest in this analysis is the P-D diagram we will compare only
the best-fitting models with LRL on the P-z plane using the much
simpler $\chi ^2$ method.

\subsection{The cosmological evolution of the progenitor population}

\subsubsection{Model A}

In the first model considered, model A, all distribution functions of
source and environmental parameters except that of the jet power,
$Q_o$, are assumed to be uniform as described by the `top-hat'
function, $f$, given in equation (\ref{tophat}). For the jet power the
exponential distribution of equation (\ref{powdis}) is assumed. The
birth function is assumed to be a power law of redshift and the
distribution of source ages is therefore given by equations
(\ref{birth1}) and (\ref{birth0}) respectively. The best-fitting model
parameters for both, the isolated galaxy and the cluster density
profile of the progenitor environment are given in Table
\ref{tab:parameter1} for $\Omega _o =1$ and Table \ref{tab:parameter0}
for $\Omega _o =0$. The resulting relative source numbers in the P-D
bins for the best-fitting model together with the value of the $\chi
^2$-test for this model are given in Table \ref{tab:sourcenumber1} for
$\Omega _o =1$ and Table \ref{tab:sourcenumber0} for $\Omega _o=0$.

\begin{table*}
{\centering
\begin{tabular}{crrrrrrrrrrrrrr}
 & \multicolumn{2}{c}{A} & \multicolumn{2}{c}{B} &
\multicolumn{2}{c}{C} & \multicolumn{2}{c}{D} & \multicolumn{2}{c}{E}
& \multicolumn{2}{c}{F} & \multicolumn{2}{c}{G} \\[-1ex] 
&
\multicolumn{2}{c}{\rule{14ex}{0.1ex}} &
\multicolumn{2}{c}{\rule{14ex}{0.1ex}} &
\multicolumn{2}{c}{\rule{14ex}{0.1ex}} &
\multicolumn{2}{c}{\rule{14ex}{0.1ex}} &
\multicolumn{2}{c}{\rule{14ex}{0.1ex}} &
\multicolumn{2}{c}{\rule{14ex}{0.1ex}} &
\multicolumn{2}{c}{\rule{14ex}{0.1ex}}\\ 
& {\em gal.} & {\em cl.} &
{\em gal.} & {\em cl.} & {\em gal.}  & {\em cl.} & {\em gal.} & {\em
cl.} & {\em gal.} & {\em cl.} & {\em gal.} & {\em cl.} & {\em gal.} &
{\em cl.}\\
\hline
$\lambda _q$ & 4 & 4 & 4 & 3 & 5 & 4 & 5 & 5 & 4 & 4 & 2 & 1 & 5 & 5\\
$n$ & 6 & 5 & 1 & 0 & 6 & 4 & 6 & 7 & 5 & 5 & 3 & 0 & 5 & 6\\
$n_d$ & \ldots & \ldots & 9 & 9 & 9 & 9 & 9 & 9 & 6 & 9 & \ldots & \ldots & \ldots & \ldots\\
$\lambda _{\beta , 0}$ & \ldots & \ldots & \ldots & \ldots & \ldots & \ldots & 4 & 1 & \ldots & \ldots & \ldots & \ldots & \ldots & \ldots\\
$n_{\beta}$ & \ldots & \ldots & \ldots & \ldots & \ldots & \ldots & 3 & 3 & \ldots & \ldots & \ldots & \ldots & \ldots & \ldots\\
$n_t$ & \ldots & \ldots & \ldots & \ldots & \ldots & \ldots & \ldots & \ldots & 1.0 & 1.0 & \ldots & \ldots & \ldots & \ldots\\
$n_{t,q}$ & \ldots & \ldots & \ldots & \ldots & \ldots & \ldots & \ldots & \ldots & \ldots & \ldots & 0.5 & 0.6 & \ldots & \ldots\\
$n_q$ & \ldots & \ldots & \ldots & \ldots & \ldots & \ldots & \ldots & \ldots & \ldots & \ldots & \ldots & \ldots & 2.0 & 2.0\\
\hline
$\chi _4^2$ & 78.1 & 171.3 & 66.3 & 58.2 & 23.0 & 26.5 & 12.8 & 24.9 & 12.0 & 11.4 & 7.7 & 17.1 & 6.9 & 12.7\\
$\sigma$ & 0\% & 0\% & 0\% & 0\% & 15\% & 7\% & 75\% & 10\% & 80\% & 83\% & 97\% & 45\% & 99\% & 75\%\\
$-\log _{10}(\rho _{tot})$ & $0.61$ & $1.87$ & $3.92$ & $5.31$ & $0.90$ & $2.88$ & $1.02$ & $2.42$ & $1.51$ & $2.17$ & $2.98$ & $5.18$ & $1.89$ & $1.71$\\
\hline
\end{tabular}}
\caption{\footnotesize The best-fitting model parameters for $\bf
\Omega _o =1$. To determine the best fir between a model and the
source distribution of the LRL sample bin 4 in Figure \ref{fig:pdbin}
was omitted. The value of the $\chi ^2$-test for this case and the
significance of the fit, $\sigma$, are given for each model. The
normalisation of the source distribution function, equation
(\ref{weight}), is given as negative logarithm of $\rho _{tot}$ in
units of Mpc$^{-3}$. As before {\em gal.} and {\em cl.} indicate the
use of the isolated galaxy and the cluster density profile
respectively.}
\label{tab:parameter1}
\end{table*}

\begin{table*}
{\centering
\begin{tabular}{crrrrrrrrrrrrrr}
 & \multicolumn{2}{c}{A} & \multicolumn{2}{c}{B} &
\multicolumn{2}{c}{C} & \multicolumn{2}{c}{D} & \multicolumn{2}{c}{E}
& \multicolumn{2}{c}{F} & \multicolumn{2}{c}{G} \\[-1ex] 
&
\multicolumn{2}{c}{\rule{14ex}{0.1ex}} &
\multicolumn{2}{c}{\rule{14ex}{0.1ex}} &
\multicolumn{2}{c}{\rule{14ex}{0.1ex}} &
\multicolumn{2}{c}{\rule{14ex}{0.1ex}} &
\multicolumn{2}{c}{\rule{14ex}{0.1ex}} &
\multicolumn{2}{c}{\rule{14ex}{0.1ex}} &
\multicolumn{2}{c}{\rule{14ex}{0.1ex}}\\ 
& {\em gal.} & {\em cl.} &
{\em gal.} & {\em cl.} & {\em gal.}  & {\em cl.} & {\em gal.} & {\em
cl.} & {\em gal.} & {\em cl.} & {\em gal.} & {\em cl.} & {\em gal.} &
{\em cl.}\\
\hline
$\lambda _q$ & 4 & 4 & 4 & 3 & 4 & 4 & 5 & 5 & 4 & 3 & 2 & 2 & 5 & 4\\
$n$ & 7 & 7 & 2 & 0 & 3 & 4 & 7 & 8 & 4 & 2 & 2 & 2 & 5 & 3\\
$n_d$ & \ldots & \ldots & 9 & 9 & 9 & 9 & 9 & 9 & 8 & 9 & \ldots & \ldots & \ldots & \ldots\\
$\lambda _{\beta , 0}$ & \ldots & \ldots & \ldots & \ldots & \ldots & \ldots & 3 & 3 & \ldots & \ldots & \ldots & \ldots & \ldots & \ldots\\
$n_{\beta}$ & \ldots & \ldots & \ldots & \ldots & \ldots & \ldots & 4 & 4 & \ldots & \ldots & \ldots & \ldots & \ldots & \ldots\\
$n_t$ & \ldots & \ldots & \ldots & \ldots & \ldots & \ldots & \ldots & \ldots & 1.0 & 1.5 & \ldots & \ldots & \ldots & \ldots\\
$n_{t,q}$ & \ldots & \ldots & \ldots & \ldots & \ldots & \ldots & \ldots & \ldots & \ldots & \ldots & 0.5 & 0.6 & \ldots & \ldots\\
$n_q$ & \ldots & \ldots & \ldots & \ldots & \ldots & \ldots & \ldots & \ldots & \ldots & \ldots & \ldots & \ldots & 2.0 & 2.0\\
\hline
$\chi _4^2$ & 44.6 & 138.3 & 25.4 & 38.3 & 13.6 & 28.5 & 12.6 & 27.5 & 10.2 & 14.4 & 7.1 & 18.1 & 8.1 & 17.6\\
$\sigma$ & 0\% & 0\% & 9\% & 0\% & 70\% & 4\% & 76\% & 5\% & 90\% & 64\% & 98\% & 38\% & 96\% & 42\%\\
$-\log _{10}(\rho _{tot})$ & 2.36 & 3.19 & 4.98 & 6.51 & 4.60 & 4.76 & 2.79 & 2.66 & 4.10 & 4.06 & 5.09 & 5.50 & 3.93 & 5.44\\
\hline
\end{tabular}}
\caption{\footnotesize The best-fitting model parameters for $\bf
\Omega _o =0$. To determine the best fir between a model and the
source distribution of the LRL sample bin 4 in Figure \ref{fig:pdbin}
was omitted. The value of the $\chi ^2$-test for this case and the
significance of the fit, $\sigma$, are given for each model. The
normalisation of the source distribution function, equation
(\ref{weight}), is given as negative logarithm of $\rho _{tot}$ in
units of Mpc$^{-3}$. As before {\em gal.} and {\em cl.} indicate the
use of the isolated galaxy and the cluster density profile
respectively.}
\label{tab:parameter0}
\end{table*}

For model A we note that the number of sources in the low luminosity
bins are comparable to the LRL sample but in all cases there are too
many powerful, large sources (bins 12 and 16). This implies that the
median hot spot advance speed in the powerful sources is too high for
the assumed set of distribution functions. The slope of the birth
functions, equations (\ref{birth1}) and (\ref{birth0}), is steep in
all cases; the exponent $n$ is in the range of 5 to 7. Although very
luminous sources are therefore, in this model, located preferentially
at a higher redshift than the less luminous ones, the increased energy
losses of the relativistic electrons in the cocoon due to the higher
energy density of the CMBR at high redshift do not sufficiently
steepen the evolutionary tracks of the most powerful sources.

\subsubsection{Model B}

The hot spot advance speed of a given source depends crucially on the
density distribution of the material the source is expanding into. An
increase in the median external density will therefore on average slow
the linear expansion of the jets. The density of the progenitor
environments at low redshift, whether they are individual galaxy or
cluster profiles, are constrained by observations. For model A most of
the objects which have linear sizes larger than those of sources in
LRL are luminous and therefore preferentially located at high
redshifts. This trend strongly suggests that the density of the
environments the progenitors are located in, increases with redshift,
and therefore we have considered models in which upper and the lower
limit of the range of the central density, $\rho _o$, varies with
redshift, according to

\begin{eqnarray}
\rho _{o,min} & = & \rho _{o,min}(z_p=0) \, \left( z_p +1 \right) ^{n_d}
\nonumber\\
\rho _{o,max} & = & \rho _{o,max}(z_p=0) \, \left( z_p +1 \right) ^{n_d}.
\end{eqnarray}

\noindent In these expressions $\rho _{o,min}(z_p=0)=5\cdot 10^{-23}$
kg/m$^3$ and $\rho _{o,max}(z_p=0)=5\cdot 10^{-21}$ kg/m$^3$ for
galactic and $\rho _{o,min}(z_p=0)=5\cdot 10^{-25}$ kg/m$^3$ and $\rho
_{o,max}(z_p=0)=5\cdot 10^{-23}$ kg/m$^3$ for cluster density profiles
as before. The parameter $n_d$ is investigated in the range 0 to
9. The best fit obtained for model B predicts a very strong evolution
of $\rho _o$ with redshift independent of the form of external density
or $\Omega _o$. If the density of the progenitor environments is
independent of other source parameters as we have assumed here, we
expect $n_d \le 3$ (see Section \ref{sec:mod}). Model B is therefore
unphysical since it requires $n_d>9$.

\subsubsection{Models C and D}

Although the exponential distribution of jet powers, equation
(\ref{powdis}), ensures that there are many more progenitors
developing weak jets as compared to those producing strong jets, the
results of model B suggest a persisting overestimate of the number of
sources with very powerful jets in the model sample. We have therefore
considered a variant of model B with a maximum jet power of $\log
_{10} (Q_{o,max}/\rm{W}) = 39.4$ in the case of $\Omega _o=1$ and
$\log _{10} (Q_{o,max}/\rm{W}) = 39.7$ for $\Omega _o =0$ instead of
$\log _{10} (Q_{o,max}/\rm{W}) = 40$. For the best-fitting model
parameters the resulting model C includes almost no sources which are
too luminous. Unless otherwise stated the following model samples are
calculated with a maximum jet power of $10^{39.4}$ W for $\Omega _o=1$
and $10^{39.7}$ W for $\Omega _o=0$.

For the determination of the best-fitting set of model parameters in
model C bin 4 was excluded from the $\chi ^2$-test since it alone
contributes up to 76\% of the total deviation of the model from the
data of the LRL sample found with the $\chi ^2$-test. The contribution
of bin 4 is so large that it dominates the fitting procedure and any
model for which the distribution of the LRL sample in the P-D plane
may be a very good representation, could be rejected as a bad fit, if
there is a strong deviation of LRL from the model sample in bin 4
alone. This suggests that some or all of the sources of the LRL sample
found in bin 4 represent a class of objects distinct from the rest of
the sample. These sources must be unusually old and/or located in very
underdense environments. In the following we will exclude bin 4 from
the $\chi ^2$-test when determining the best-fitting model parameters.

Although model C represents an improvement over models A and B,
particularly in the case of galactic density profiles and $\Omega
_o=0$, the predicted redshift evolution of the central density, $\rho
_o$, is again steeper than the physical limit in all cases. Another
possible mechanism which could reduce the linear size of powerful jets
at high redshift is the predominance of flat density profiles for the
environments of progenitors at high $z$. This is consistent with the
scenario of the cosmological evolution of the environments of the
progenitors outlined above since we expect that at early stages of the
evolution of regions which have decoupled from the Hubble flow, the
density gradient within these regions is small. In model D we
therefore introduce an exponential distribution for $\beta$ peaking at
$\beta =1$ which steepens for increasing redshift;

\begin{eqnarray}
N_5 (\beta) \, d\beta & = & \frac{\lambda _{\beta} e^{-\lambda
_{\beta} \left( \beta -\beta_{min} \right)}}{1-e^{-\lambda _{\beta}
\left( \beta _{max}-\beta _{min}\right)}} \, d\beta \nonumber\\
\lambda _{\beta} & = & \lambda _{\beta , 0} + \left( z_p +1 \right)
^{n_{\beta}} -1,
\end{eqnarray}

\noindent where $\lambda _{\beta,0}$ and $n_{\beta}$ vary from 0 to 4
in steps of 1. Although model D improves the model fit significantly
in the case of individual galaxy density profiles, the predicted
evolution of the central density is still too steep with $n_d =9$.

\subsubsection{Model E}

For models A to D we have assumed that the maximum age, $t_{max}$, of
all radio sources is $10^9$ years. If $t_{max}$ decreases with
increasing redshift, we expect the median linear size of radio sources
to be smaller at higher redshift. Because of the strong correlation of
redshift and radio luminosity in flux limited samples this may imply a
decreasing median linear size with increasing luminosity. To
investigate this we introduce

\begin{equation}
t_{max} = t_{max,0} \left( z_p +1 \right) ^{-n_t}
\end{equation}

\noindent in model E, where $t_{max,0}$ is $10^9$ years and
$n_t$ varies from 0.0 to 4.0 in steps of 0.5. Although the fit in
model E is improved as compared to the previous models, the predicted
evolution of the central density, $\rho _o$, is again very steep. The
decrease in the maximum life time of sources at high redshift which
feature preferentially powerful jets, alone is not sufficient to
reduce the median linear size of these objects as required by the
observations. Note that, for the assumption of cluster density
profiles, model E of all model samples calculated here, provides the
best fit in both cosmologies with the observed LRL sample.

\subsubsection{Models F and G} 

To make progress and obtain a model which fits the distribution of the
LRL sample in the P-D plane without invoking an unphysical evolution
of the density of the progenitor environment, we relax the assumption
that all jet and environment parameter are independent of each other.

The model of the evolution of the radio luminosity of powerful radio
galaxies described by KDA assumes that the power of the jet,
$Q_o$, is constant over the entire life time of the source. Together
with the uniform maximum age for these objects assumed here, this
implies that the total energy transported from the AGN to the large
scale structure, $E_{tot}$, during $t_{max}$ is linearly proportional
to $Q_o$, which may not necessarily be the case in radio galaxies. If
the powerful jets exhaust their energy supply faster than their weaker
counterparts, e.g. $t_{max} \propto Q_o^{-n_{t,q}}$, than this linear
dependence is replaced by $E_{tot} \propto Q_o^{1-n_{t,q}}$ and we
expect the median linear size of the powerful sources to decrease. For
model F we take

\begin{equation}
t_{max} = t_{max,0} \left( \frac{Q_o}{Q_{o,min}} \right) ^{-n_{t,q}},
\end{equation}

\noindent where $n_{t,q}$ is varying from 0.0 to 0.9 in steps of
0.1. We also take the maximum jet power, $Q_{o,max}$, to be
$10^{39.7}$ W for $\Omega _o =1$ and $10^{40}$ W for $\Omega _o=0$
which improves the fit at the highest luminosities (bins 14 and
15). For the assumption of individual galaxy density profiles the
model fit improves to more than 97\% significance for both cosmologies
and the model predicts $E_{tot} \propto Q_o^{0.5}$. For cluster
density profiles the fit is far less good.

Rawlings \& Saunders (1991)\nocite{rs91} find that the power of the
jet is roughly comparable to the Eddington luminosity of the black
hole in the AGN powering the jet at $z\sim 1$. Kormendy \& Richstone
(1995)\nocite{kr95} find a correlation of the mass of the central
black hole and the mass of the bulge in nearby galaxies. Since we
model the density profile of the progenitor environment with a power
law (equation \ref{king}), the total mass of the progenitor is
proportional to the central density of this distribution, $\rho
_o$. This may imply that the power of the jet in radio galaxies
depends on the mass and the density of gas in the progenitor
object. In model G we therefore take

\begin{eqnarray}
\rho _{o,min} & = & \rho _{o,min} (z_p=0) \, \left(
\frac{Q_o}{Q_{o,min}} \right) ^{n_q} \nonumber\\
\rho _{o,max} & = & \rho _{o,max} (z_p=0) \, \left(
\frac{Q_o}{Q_{o,min}} \right) ^{n_q}, 
\end{eqnarray}

\noindent where $\rho _{o,min}(z_p=0)=5\cdot 10^{-23}$ kg/m$^3$ and
$\rho _{o,max}(z_p=0)=5\cdot 10^{-21}$ kg/m$^3$ for galactic and $\rho
_{o,min}(z_p=0)=5\cdot 10^{-25}$ kg/m$^3$ and $\rho
_{o,max}(z_p=0)=5\cdot 10^{-23}$ kg/m$^3$ for cluster density profiles
as before. The maximum age of all sources is assumed to be $10^9$
years and the maximum jet power is set to $10^{39.4}$ W for $\Omega
_o=1$ and $10^{39.7}$ W for $\Omega _o=0$ as before. The exponent
$n_q$ is varied from 0.0 to 4.5 in steps of 0.5. The significance of
the fit for the assumption of galactic density profiles is comparable
to that of model F, while for cluster density profiles the fit
improves compared to the last model.

\begin{table*}
{\centering
\begin{tabular}{ccrrrr@{\hspace{10ex}}crrrr}
 & \multicolumn{5}{c@{\hspace{10ex}}}{$\Omega _o=1$} & \multicolumn{5}{c}{$\Omega _o=0$}
\\[-1ex] 
& \multicolumn{5}{c@{\hspace{10ex}}}{\rule{42ex}{0.1ex}} & 
\multicolumn{5}{c}{\rule{41ex}{0.1ex}}\\
& & \multicolumn{2}{c}{F} & \multicolumn{2}{c@{\hspace{10ex}}}{G} & & \multicolumn{2}{c}{F} & \multicolumn{2}{c}{G}\\[-1ex]
& & \multicolumn{2}{c}{\rule{14ex}{0.1ex}} & \multicolumn{2}{c@{\hspace{10ex}}}{\rule{15ex}{0.1ex}} & & \multicolumn{2}{c}{\rule{14ex}{0.1ex}} & \multicolumn{2}{c}{\rule{15ex}{0.1ex}}\\
bin & LRL & {\em gal.} & {\em cl.} & {\em gal.} & {\em cl.} & LRL & {\em gal.} & {\em cl.} & {\em gal.} & {\em cl.}\\
\hline
1 & 6 & 7.09 & 4.44 & 7.46 & 6.06 & 6 & 8.96 & 9.41 & 8.56 & 6.14\\
2 & 5 & 1.76 & 2.44 & 1.98 & 2.67 & 5 & 2.29 & 3.81 & 2.35 & 3.42\\
3 & 6 & 6.92 & 8.02 & 8.93 & 10.66 & 5 & 7.59 & 10.92 & 9.39 & 11.36\\
4 & 5 & 3.31 & 5.27 & 3.64 & 3.47 & 5 & 4.20 & 5.57 & 4.53 & 5.99\\
5 & 23 & 17.90 & 20.15 & 20.83 & 19.95 & 20 & 15.16 & 17.31 & 17.52 & 19.59\\
6 & 3 & 5.56 & 8.59 & 4.76 & 3.59 & 3 & 7.46 & 7.38 & 6.71 & 8.05\\
7 & 35 & 34.36 & 31.62 & 31.91 & 28.49 & 32 & 29.22 & 26.72 & 29.22 & 28.03\\
8 & 3 & 6.04 & 10.09 & 4.78 & 3.48 & 5 & 7.78 & 6.82 & 5.65 & 6.62\\
9 & 9 & 11.01 & 12.96 & 12.73 & 11.13 & 14 & 15.13 & 13.14 & 15.43 & 15.00\\
10 & 39 & 39.82 & 30.45 & 42.39 & 50.54 & 34 & 31.26 & 28.96 & 37.34 & 33.11\\
11 & 1 & 0.04 & 0.00 & 0.00 & 0.00 & 1 & 0.82 & 1.13 & 0.13 & 0.08\\
12 & 0 & 0.19 & 0.00 & 0.02 & 0.00 & 3 & 3.78 & 4.82 & 1.22 & 0.47\\
13 & 5 & 0.28 & 0.00 & 0.03 & 0.00 & 7 & 0.76 & 0.94 & 0.12 & 0.01\\
14 & 0 & 0.00 & 0.00 & 0.00 & 0.00 & 0 & 0.00 & 0.00 & 0.00 & 0.00\\
15 & 0 & 0.00 & 0.00 & 0.00 & 0.00 & 0 & 0.00 & 0.00 & 0.00 & 0.00\\
16 & 0 & 1.69 & 4.47 & 1.06 & 0.69 & 0 & 2.22 & 2.24 & 1.21 & 1.92\\
17 & 0 & 0.00 & 0.00 & 0.00 & 0.00 & 0 & 0.00 & 0.00 & 0.00 & 0.00\\
\hline
$\chi ^2$ & \ldots & 118.7 & $\infty$ & 1186.8 & $\infty$ & \ldots & 64.7 & 52.5 & 400.2 & 7178.6\\
$\chi _r^2$ & \ldots & 13.4 & 21.23 & 10.6 & 10.6 & \ldots & 13.3 & 12.6 & 10.2 & 10.6\\
$\sigma$ & \ldots & 50\% & 10\% & 72\% & 72\% & \ldots & 51\% & 56\% & 74\% & 72\%\\
\hline
\end{tabular}}
\caption{\footnotesize The number of sources in each P-z bin as
defined in Figure \ref{fig:pzbin} for the LRL sample and models F and
G for $\Omega _o =1$ on the left and $\Omega _o =0$ on the right. {\em
gal.} and {\em cl.} indicate the use of the isolated galaxy and the
cluster density profile respectively. $\chi ^2$ is the result of the
complete $\chi ^2$-test for the respective model and $\chi _r ^2$ is
the result of the restricted $\chi ^2$-test which omits bins 11, 12
and 13. $\sigma$ is the significance of the fit using the restricted
$\chi ^2$-test.}
\label{tab:zcomp}
\end{table*}

\subsubsection{Comparison of models}

Models F and G together with the assumption of galactic density
profiles for the environment of the progenitors represent the best fit
with the source distribution of the LRL sample in the P-D plane in
both cosmologies of all the model calculations presented here. For the
assumption of cluster density profiles model E provides a better fit
than either model F or G. However, the very steep evolution of the
central density, $\rho _o$, with redshift predicted by model E is
unphysical and we will therefore restrict attention in the following
section to models F and G with the assumption of galactic density
profiles for the environments of the progenitor objects. The fact that
the best-fitting parameters for models F and G are very similar, if
not completely identical, for the two cosmologies investigated is
remarkable given the higher radio luminosity required of sources to be
included in the model sample and the higher maximum jet power used in
the case of $\Omega _o=0$. The source with the highest redshift in the
LRL sample, 3C9, is located at roughly $z=2.1$. The derived linear
size and radio luminosity for an observed source at $z=2.1$ with
measured angular size and radio flux density are a factor 1.6 and 2.6
respectively greater for $\Omega _o=0$ than for $\Omega _o=1$. The
resulting difference in the source distribution of the LRL sample on
the P-D plane can apparently be accounted for in model F and G simply
by the inclusion of sources with higher jet power in the model sample
for the case $\Omega _o=0$, without changing the model parameters.

To determine which of the two models is in better agreement with the
LRL sample we now compare the source distribution of the model samples
with the best-fitting model parameters as given in Tables
\ref{tab:parameter1} and \ref{tab:parameter0} in the P-z plane with
that of the LRL sample. The results of the $\chi ^2$-test for the
binning of the P-z plane as introduced in Figure \ref{fig:pzbin} are
summarised in Table \ref{tab:zcomp}. Here again we have omitted the
three sources of the LRL sample in bin 4 of the binning of the P-D
plane. Note here, that the low value for the maximum jet power used in
models F and G implies effectively a maximum radio luminosity for
sources in the model samples. Together with the flux limit this means
that the agreement of the model samples with the LRL sample in the P-z
plane will be poor at the highest redshifts and also for the highest
luminosities. If we restrict attention to lower redshifts and lower
radio luminosities by omitting bins 11, 12 and 13 from the $\chi
^2$-test, we find good agreement between the model samples and
LRL. See also the following section for further discussion of this
point. In both cosmologies the agreement between model G and the LRL
sample is better than that between model F and LRL.

\section{The FRII luminosity function}

Using the results found in the previous section for the source
distribution function, equation (\ref{weight}), it is now possible to
derive the luminosity function of FRII sources, $\rho
(P_{\nu},z)$. Following the definition of Peacock (1985)\nocite{jp85},
$\rho (P_{\nu},z)$ is the number of radio sources per comoving volume
measured in Mpc$^3$ and per unit interval of $\log _{10} (P_{\nu})$
with $P_{\nu}$ measured in W Hz$^{-1}$ sr$^{-1}$. This luminosity
function is identical to the source distribution function divided by
$\log _{10} (P_{\nu})$ but it is a function of the radio luminosity,
$P_{\nu}$ at frequency $\nu$ and redshift, instead of the source and
environmental parameters. To be able to compare our results with those
of Dunlop \& Peacock (1990)\nocite{dp90} we use their observing
frequency 2.7 GHz and calculate the `observed' spectral index between
2.7 GHz and 1.4 GHz. The luminosity function is then determined for 10
bins of equal width in $\log _{10} P_{2.7}$ from 23.0 to 28.0 in the
same way as the model samples in the previous section. Note, that we
are now interested in all radio sources in the whole universe and not
only in those which are observable to us. We therefore do not have to
multiply equation (\ref{weight}) by equation (\ref{vol1}) or
(\ref{vol0}) respectively.

\begin{figure*}
\centerline{\epsfig{file=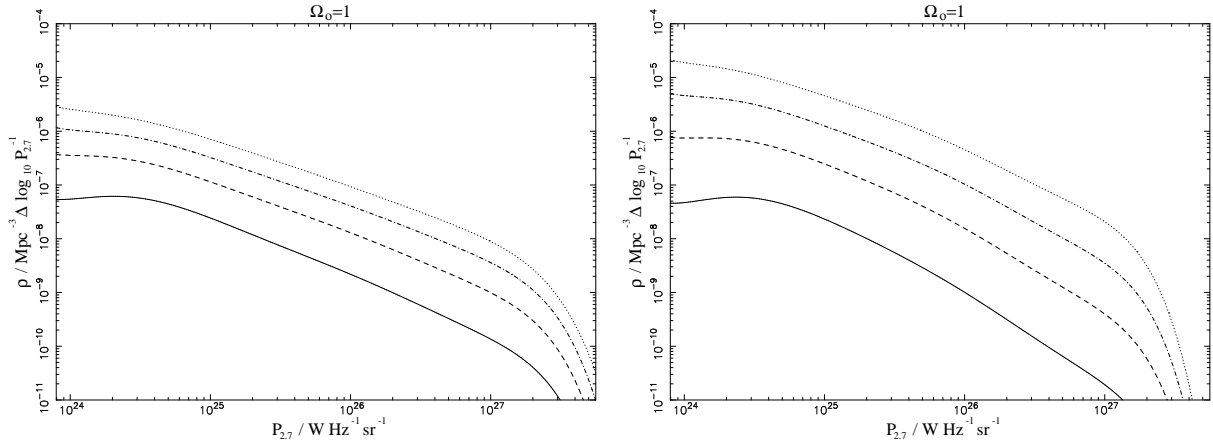, height=16cm, angle=270}}
\caption{\footnotesize The FRII luminosity function for $\bf \Omega _o
=1$. Left: model F, right: model G. Solid line: z=0, dashed: z=0.5,
dot--dashed: z=1.0 and dotted: z=1.5.}
\label{fig:lum1}
\end{figure*}

\begin{figure*}
\centerline{\epsfig{file=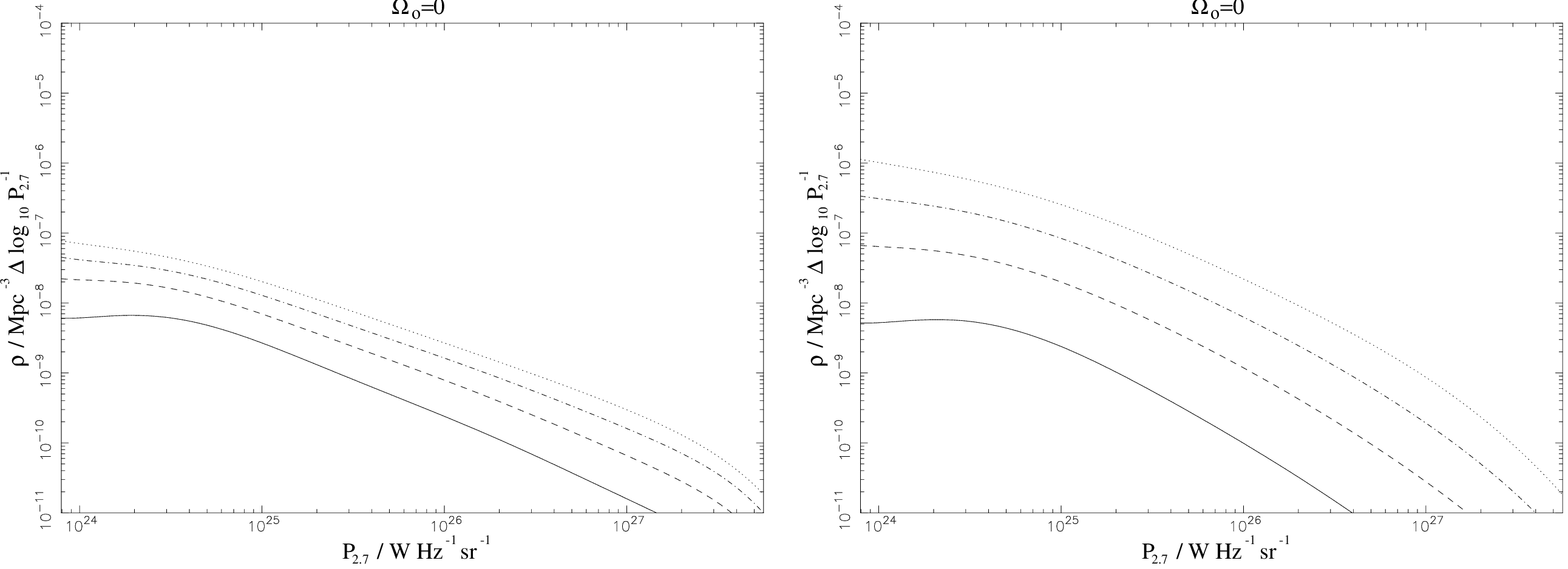, height=16cm, angle=270}}
\caption{\footnotesize The FRII luminosity function for $\bf \Omega _o
=0$. Left: model F, right: model G. Solid line: z=0, dashed: z=0.5,
dot--dashed: z=1.0 and dotted: z=1.5.}
\label{fig:lum0}
\end{figure*}

Figures \ref{fig:lum1} and \ref{fig:lum0} show the resulting luminosity
function for models F and G with the assumption of galactic density
profiles in the environments of the progenitors. The flattening of the
function for the lowest luminosities and the steepening at the high
luminosity end are caused by the limits placed on the jet power,
$Q_o$, and the density of the progenitor environment. Sources of lower
luminosity than approximately $P_{2.7} = 2\cdot 10 ^{24}$ W Hz$^{-1}$
sr$^{-1}$ are likely to have low jet powers which makes their jets
susceptible to turbulence. Since FRI sources are not included in the
model, the space density of low luminosity sources predicted here is a
lower limit.

Almost all of the sources in the LRL sample are close to the
observational cut-off introduced by the flux limit of the sample (see
Figure \ref{fig:pzbin}). The one source in LRL in bin 11, the quasar
3C196, and the most distant source in LRL, the quasar 3C9, in bin 13
are clearly peculiar exceptions to this observation since they are far
too luminous for their respective redshifts. The remaining sources in
the LRL sample in bins 12 and 13 are difficult to reconcile with the
very low relative source numbers in the model samples for models F and
G, particularly in the case of $\Omega _o=1$.

\begin{figure*}
\centerline{\epsfig{file=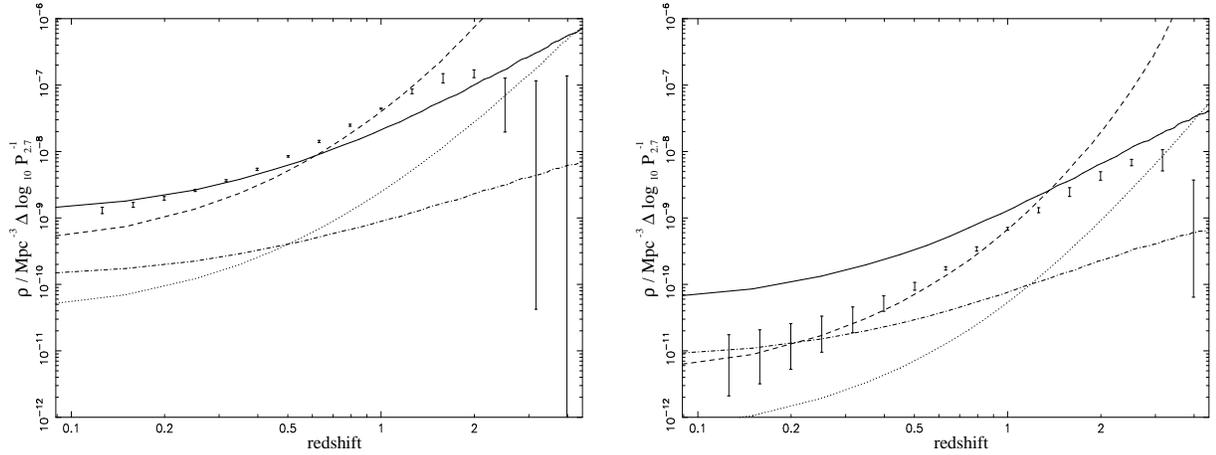, height=16cm, angle=270}}
\caption{\footnotesize The FRII luminosity function as a function of
redshift. Left: $P_{2.7} =10^{26}$ W Hz$^{-1}$ sr$^{-1}$, right:
$P_{2.7} =10^{27}$ W Hz$^{-1}$ sr$^{-1}$. Solid line: model F for
$\Omega _o=1$, dashed: model G for $\Omega _o=1$, dot--dashed: model F
for $\Omega _o=0$ and dotted: model G for $\Omega _o=0$. The error
bars show the range of the luminosity function for $\Omega _o=1$ as
predicted by the free-modelling approach of Dunlop \& Peacock (1990).}
\label{fig:lumc}
\end{figure*}

By introducing rather low maximum jet powers in the previous section
it was possible to reduce the number of sources of high luminosity in
the model samples but this also leads to a lack of sources in the
model samples with luminosities and redshifts comparable to the
highest luminosities and redshifts in the LRL sample. When included,
the number of sources with the highest jet powers at high redshift is
overpredicted by the model. This implies that, although the simple
power law assumed for the birth function of radio sources, equations
(\ref{birth1}) and (\ref{birth0}), provides good model fits at low
redshift it is not a good fit to the `true' birth function at high
redshift. The `true' birth function must flatten or even turn over at
high redshift to explain the comparatively small number of sources at
these cosmological epochs. This is supported by a comparison of our
results with those of Dunlop \& Peacock (1990)\nocite{dp90} presented
in Figure \ref{fig:lumc}. The flattening and the turn-over of the
luminosity function predicted by their free-modelling approach occurs
at redshifts comparable to those at which our model is overpredicting
the number of sources with powerful jets when these are included in
the model samples.

Dunlop \& Peacock (1990)\nocite{dp90} used an extensive data base of
observed sources with a flux limit at their observing frequency at 2.7
GHz of two orders of magnitude lower than that of LRL at 178 MHz as
used in the previous section to constrain and normalise the model. The
agreement between their results and the prediction of our models,
particularly model G, is therefore remarkable.

The form of the cosmological evolution of the radio source population
can be dominated either by space density evolution, or by luminosity
evolution. The evolution of the FRII population as predicted by models
F and G is dominated by the space density evolution of radio
sources. This is a result of our assumption of independence of source
and environment parameters. The ratio of sources with high jet power
and those with low jet power is the same at each redshift and only the
total number of radio sources is a function of z. Objects of similar
jet power are also located in similar environments and have therefore
similar radio luminosities regardless of their redshift. Exceptions to
this rule are large sources which are already affected by the inverse
Compton losses of the relativistic electrons in their cocoons. These
sources have lower luminosities at higher redshifts because of the
higher energy density of the CMBR at these epochs and show therefore
some luminosity evolution. In the case of pure density evolution or
pure luminosity evolution the slope of the luminosity function will
stay constant at all redshifts. Measuring the slope of the luminosity
function predicted by our models between $P_{2.7} =10 ^{25}$ W
Hz$^{-1}$ sr$^{-1}$ and $P_{2.7} =10 ^{27}$ W Hz$^{-1}$ sr$^{-1}$ in
Figures \ref{fig:lum1} and \ref{fig:lum0} we find that this slope
decreases from $-1.2$ at $z=0$ to $-0.9$ at $z=1.5$ in the case of
model F. For model G the slope decreases from $-1.5$ at $z=0.0$ to
$-1.2$ at $z=1.5$. This slight flattening of the luminosity function
with increasing redshift was also found by Dunlop \& Peacock
(1990)\nocite{dp90} when they used a combined density-luminosity
evolution model to fit their data. 

Despite the luminosity evolution of large sources predicted by our
models the overall evolution of the FRII population will be dominated
by density evolution because of the set-up of our models. Fainter
samples of FRII sources indicate that the evolution of the radio
luminosity function for sources of lower luminosity is different from
that of high luminosity sources (e.g. Dunlop \& Peacock
1990\nocite{dp90}). Some form of luminosity evolution therefore must
take place. To model this behaviour of the radio luminosity function
we would have to introduce a dependence of the birth function on the
jet power. This may also make it possible to avoid the low maximum jet
power imposed on models F and G. However, this dependence would lead
to an increase in the number of model parameters and since the LRL
sample includes only the most luminous radio sources at any given
redshift, the constraints on these additional parameters would be
weak. We have therefore not investigated such models and our results
presented here apply mostly to the high luminosity end of the radio
luminosity function.

\section{Linear size evolution}

In order to determine the positions of a radio sources with given jet
and environment parameters in the P-D plane we have already calculated
the linear sizes of these sources. It is therefore straightforward to
determine the median length of radio sources, $D_{med}$, as a function
of redshift.

If fitted by a power law, $D_{med} \propto \left( 1+z \right) ^{n_D}$,
we find $n_D = -1.3$ for model F and $n_D =-1.1$ for model G in both
cosmologies. The model data is fitted well by a power law (correlation
greater than 0.97) to $z=1.5$. For higher redshifts the small number
of sources in the model samples effectively prevents the determination
of a meaningful value of $D_{med}$. Both values for $n_D$ are within
the error range of the value found by Neeser {\em et al.}
(1995)\nocite{ner95} from a statistical analysis of the sources in the
LRL and the 6C sample (e.g. Eales 1985\nocite{se85}). There are no
FRI-type objects in our model samples and our result can therefore be
taken as a confirmation of the large selection effect pointed out by
Neeser {\em et al.}  (1995)\nocite{ner95} in the derivation of the
linear size evolution of radio sources.

When considering all radio sources in the universe we find $n_D= -0.4$
for both models in both cosmologies. Also, the median linear size at
$z=0$ is a factor 5 greater when all sources are taken into account as
compared to the median linear size of the sources in the flux-limited
model sample. This is consistent with larger sources being excluded
from a flux limited sample because they are less luminous. The steeper
linear size evolution in a flux-limited sample shows that this bias
increases with increasing redshift.

\section{Discussion}

In the models considered in this paper we have assumed the birth
function, i.e. the fraction of the radio source population becoming
active as a function of cosmological epoch, to increase monotonically
with redshift. Since the central black hole powering AGNs must form at
some cosmological epoch, one expects the birth function to turn over
at some redshift and our model will predict too many sources at higher
redshifts. This effect is indeed present in our modelling and led us
to introduce a lower maximum jet power since only sources with the
highest jet powers and therefore highest radio luminosities are
observable beyond the turn-over redshift. We find that our model fits
the observational data well out to a redshift of about 1.5. The
turn-over in the birth function, and therefore the peak in the radio
luminosity function, must therefore occur beyond, but not far from,
$z=1.5$. This is consistent with the results of Dunlop \& Peacock
(1990)\nocite{dp90}.

The FRII objects in the LRL sample represent the brightest radio
sources in the universe at their respective cosmological epoch. They
must therefore contain the most powerful AGNs and their cosmological
evolution puts important constraints on the formation of structure in
the universe. The median value of the jet power of sources in our
model samples with the same flux limit as LRL as a function of
redshift is well fitted by a power law and we find

\begin{equation}
Q_{o,med} = \left\{
\begin{array}{ll}
10^{38.0} \, \left( 1+z \right) ^{5.0} & \mbox{W ; model F} \\
10^{37.6} \, \left( 1+z \right) ^{5.4} & \mbox{W ; model G}
\end{array}
,\right.
\label{qmed1}
\end{equation}

\noindent for $\Omega _o =1$ and

\begin{equation}
Q_{o,med} = \left\{
\begin{array}{ll}
10^{38.0} \, \left( 1+z \right) ^{5.9} & \mbox{W ; model F} \\
10^{37.5} \, \left( 1+z \right) ^{6.2} & \mbox{W ; model G}
\end{array}
,\right.
\label{qmed0}
\end{equation}

\noindent for $\Omega _o =0$. The boundaries of the range of central
densities of the progenitor environment, $\rho _o$, are constant for
model F while they are proportional to $Q_o^2$ for model G. We
therefore expect the median value of the central density as a function
of redshift in the model samples to be constant in model F and to
increase with redshift as determined by equations (\ref{qmed1}) and
(\ref{qmed0}) in model G. However, the introduction of a flux limit
may result in a deviation from this and indeed we find

\begin{equation}
\rho _{o,med} = \left\{
\begin{array}{ll}
10^{-21.0} \, \left( 1+z \right) ^{0.5} & \mbox{kg/m$^3$ ; model F}\\
10^{-19.8} \, \left( 1+z \right) ^{10.2} & \mbox{kg/m$^3$ ; model G}
\end{array}
,\right.
\label{rmed1}
\end{equation}

\noindent for $\Omega _o=1$ and

\begin{equation}
\rho _{o,med} = \left\{
\begin{array}{ll}
10^{-21.0} \, \left( 1+z \right) ^{0.4} & \mbox{kg/m$^3$ ; model F}\\
10^{-19.9} \, \left( 1+z \right) ^{11.8} & \mbox{kg/m$^3$ ; model G}
\end{array}
,\right.
\label{rmed0}
\end{equation}

\noindent in the case of $\Omega _o=0$. 

Equations (\ref{rmed1}) and (\ref{rmed0}) imply that the total mass of
the progenitors of the most luminous radio galaxies in model G
increase quickly with increasing redshift. Assuming that the gas in
these objects is distributed according to equation (\ref{king}) with
$\beta =2$, the total mass of the gas in the progenitor out to a
radius $R_x$ is given by

\begin{equation}
M_{tot} = \pi \, \rho _o \, a_o^3 \, \left( \frac{R_x}{a_o} - \arctan
\frac{R_x}{a_o} \right).
\label{mtot}
\end{equation}

\noindent For the observed galactic density profiles, which we have
used in this analysis, Canizares {\em et al.} (1987)\nocite{cft87} use
$R_x =50 a_o =50$ kpc for the largest radius at which X-ray emission
is detected in their objects. In both cosmologies, we derive total
masses for the progenitors within this radius of order $10^9$
M$_{\odot}$ for model F and of order $10^{10}$ M$_{\odot}$ for model G
at $z=0$. Out to redshift 1.5 this mass increases insignificantly in
model F for both cosmologies. However, in model G we find $M_{tot}$ at
$z=1.5$ to be of order $10^{14}$ M$_{\odot}$ for $\Omega _o=1$ and of
order $10^{15}$ M$_{\odot}$ for $\Omega _o=0$. These derived mass
concentrations for the radio source progenitors within a radius of 50
kpc are much higher than in any observed object. This may rule out
model G.

Rawlings \& Saunders (1991)\nocite{rs91} showed that the radiated
luminosity of AGNs is roughly equal to the power of the jets
associated with them. Since at least the radiated energy of AGNs is
provided by accretion of material onto a central black hole, this
implies that the jet power can be written as

\begin{equation}
Q_o = \epsilon \, L_{Ed} = 1.25 \cdot 10^{31} \, \epsilon \,
\frac{M_{BH}}{M_{\odot}} \, \rm{W},
\label{edd}
\end{equation}

\noindent where $L_{Ed}$ is the Eddington luminosity of the central
black hole with mass $M_{BH}$. $\epsilon$ is the fraction of the
Eddington luminosity at which the AGN is supplying the jet with
energy. In order for accretion onto the central black hole to be
possible, $\epsilon \le 1$ is required. Theories of the formation of
structure in the universe predict that the mass of the central black
hole in any concentration of matter is proportional to the mass of the
entire object, $M_{tot}$ (e.g. Efstathiou \& Rees
1988\nocite{er88}). This is confirmed by the masses of central black
holes derived from observations of nearby galaxies (Kormendy \&
Richstone 1995\nocite{kr95}). With equation (\ref{mtot}) this implies
$Q_o \propto \epsilon \rho _o$. Using equations (\ref{qmed1}),
(\ref{qmed0}), (\ref{rmed1}), and (\ref{rmed0}) we find the following
expressions for $\epsilon$ of the brightest radio galaxies as a
function of redshift.

\begin{equation}
\epsilon \propto \left\{
\begin{array}{ll}
\left( 1+z \right)^{4.5} & \mbox{; $\Omega _o =1$, model F}\\
\left( 1+z \right)^{-4.9} & \mbox{; $\Omega _o =1$, model G}\\ 
\left( 1+z \right)^{5.5} & \mbox{; $\Omega _o =0$, model F}\\
\left( 1+z \right)^{-5.6} & \mbox{; $\Omega _o =0$, model G}
\end{array}
\right.
\label{eff}
\end{equation}

\noindent For increasing redshift $\epsilon$ is increasing for model F
while it is decreasing for model G. Rawlings \& Saunders
(1991)\nocite{rs91} find the highest jet powers ($\sim 10^{40}$ W) in
sources at redshift 1. If $\epsilon =1$ in these sources, i.e. they
are fuelling their jets at the Eddington limit, at $z=1.5$, then the
mass of their central black holes is of order $10^9$ M$_{\odot}$. In
this case we find $\epsilon$ at $z=0$ for model F to be equal to 0.016
for $\Omega _o=1$ and $\epsilon =0.006$ if $\Omega _o =0$. For model G
we find the rate at which energy is supplied to the jet to be highly
super-Eddington at redshift 0 ($\epsilon =85$ for $\Omega _o=1$ and
$\epsilon =162$ for $\Omega _o=0$). If $\epsilon =1$ in the brightest
radio galaxies at $z=0$ instead of at $z=1.5$, then model G predicts
AGNs with black holes in their centre which are sub-Eddington for all
redshifts. However, this implies that the mass of the central black
hole in the most powerful sources at $z=1.5$ is equal to several
$10^{10}$ solar masses or higher. Although the existence of black
holes with such large masses at high redshift can not be completely
ruled out, the number of objects hosting such massive black holes is
certainly very small (e.g. Efstathiou \& Rees 1988). Note also, that
in model F $\epsilon \propto Q_o$ while in model G $\epsilon \propto
Q_o^{-1}$. This implies that the energy conversion mechanism driving
the jets in radio galaxies is more efficient in sources with high jet
powers then in those with weaker jets in model F while in model G
exactly the opposite is true.

In model F the life time of radio sources depends on the power of the
jets. The median life time of the brightest radio sources is therefore
also a function of redshift and we find $t_{max} = 3.2 \cdot 10^7$
years for $\Omega _o=1$ and $t_{max} =2.2 \cdot 10^7$ years in the
case of $\Omega _o=0$ at $z=1.5$. This is consistent with observed
spectral ages which usually do not exceed a few $10^7$ years
(Alexander \& Leahy 1987\nocite{al87}). If the life time of the jets
in radio galaxies is limited by the supply of fuel to their AGNs, then
this implies that powerful jets are not only more efficient than
weaker jets but also that they exhaust their fuel supply faster.

The very high mass concentrations in the environment of the
progenitors and the need for black holes of extremely high mass in the
centre of the AGNs driving the most powerful jets in model G, suggest
that this model is unphysical. However, KA point out that the general
dynamics of a radio source with given jet and environment parameters
does not depend on $\rho _o$ and $a_o$ separately but only on the
combination $\rho _o a_o^{\beta}$ of these two quantities. The same is
true for the radio luminosity of the source derived from the model of
KDA. The only difference between the evolutionary tracks through the
P-D diagram of two sources with the same value of $\rho _o
a_o^{\beta}$ but different core radii and central densities is caused
by the different fractions of their life time they spend in the three
density regimes introduced in Section 2.2. In the following we will
neglect the comparatively small effect these differences will have on
the model parameters of model G and replace $\rho _o$ by $\rho _o
a_o^{\beta}$. With this we obtain $\rho _o a_o^{\beta} \propto Q_o^2$
for model G. For the assumption that $\epsilon =1$ for all values of
$Q_o$ we then find $a_o \propto Q_o^{1/(3-\beta)}$ and $\rho _o
\propto Q_o^{(6-3\beta)/(3-\beta)}$. Using the redshift dependence of
the jet power of the brightest radio galaxies, equations (\ref{qmed1})
and (\ref{qmed0}), and assuming $\beta=2$ we note that $\rho _o$ is
now a constant while $a_o \propto (1+z)^{5.4}$ for $\Omega _o =1$ and
$a_o \propto (1+z)^{6.2}$ for $\Omega _o=0$. Taking $\rho _o = 5\cdot
10^{-22}$ kg/m$^3$, which is the median value of the distribution of
$\rho _o$ for the galactic density profiles, and $a_o =1$ kpc at
$z=0$, we find from equation (\ref{mtot}) $M_{tot} \sim 10^{12}$
M$_{\odot}$ within a radius of 50 kpc at $z=1.5$ for both
cosmologies. This is comparable to the mass contained within the same
radius in M87 found from globular cluster dynamics (Cohen \& Ryzhov
1997\nocite{cr97}) and from thermal X-ray emission (Nulsen \&
B\"{o}hringer 1995\nocite{nb95}). Note, however, that in the case
discussed here, the core radius of the gas distribution of the
environment, $a_o$, increases to 141 kpc for $\Omega _o =1$ and 293
kpc for $\Omega _o=0$ at redshift $z=1.5$. Both values are much larger
than those expected for the gas density profiles of individual
galaxies and are more consistent with the gas distributions in galaxy
clusters (Jones \& Forman 1984\nocite{jf84}). If $\epsilon$ for the
most luminous radio galaxies is not constant but decreasing with
decreasing jet power, as is the case in model F, $a_o$ will increase
even more strongly with redshift while $\rho _o$ may decrease with
increasing redshift making the distribution of the gas in the
environments of radio galaxies at high redshift even more similar to
that in galaxy clusters.

Recently Best {\em et al.} (1998)\nocite{blr98} showed that the host
galaxies of the most luminous radio galaxies at $z\sim 1$ are massive,
highly evolved systems, presumably similar to the progenitors of the
Brightest Cluster Galaxies (BCG). This is consistent with the tendency
of the most luminous radio galaxies at high redshift to be located in
richer environments than their low redshift counterparts (e.g. Hill \&
Lilly 1991\nocite{hl91}). If we allow for the evolution of the core
radius, $a_o$, of the density distribution in the environment of radio
galaxies, our model G predicts a similar change in the environments of
the most luminous radio galaxies from small, dense gas halos of
individual galaxies at low redshift to more extended structures
reminiscent of present day galaxy clusters at high redshift. This also
explains the poor fit of our models with the LRL sample for the
assumption of cluster-like density profiles in their environments at
all redshifts.

The strong evolution of the radio luminosity of the most luminous
radio galaxies in the universe with cosmic epoch (see Figure
\ref{fig:pzbin}) is explained differently in models F and G. For the
assumption that the mass of the central black hole is proportional to
the total mass of the radio source progenitor, model F predicts that
the mass of the central black hole powering the most luminous radio
sources is essential constant with redshift. The strong decrease in
radio luminosity is therefore caused by a less and less efficient
energy conversion process which expresses itself in the decrease of
$\epsilon$ with cosmic time. We have shown that model G is consistent
with a constant value of $\epsilon$ and the decrease of the radio
luminosity is then in this model caused by a decrease of the mass of
the central black hole in the centre of the most luminous radio
galaxies. In the latter scenario some additional process is needed to
explain why the most massive black holes do not produce jets with high
jet powers at low redshift. The faster virialisation of the material
in large objects like galaxy clusters as opposed to smaller groups
could prevent gas from reaching the centre of potential radio source
hosts within these rich environments at low redshift and thereby
depriving the black holes in these objects of fuel (i.e. Best {\em et
al.} 1998\nocite{blr98}, Ellingson {\em et al.}
1991\nocite{egy91}). This may explain why radio sources in clusters at
low redshift are usually of type FRI which implies that their jets are
comparatively weak.

\section{Conclusions}

Based on the models for the evolution of the linear size and radio
luminosity of powerful extragalactic radio sources of type FRII
presented in KA and KDA, we have investigated the distributions
of various jet and environment parameters and their evolution with
redshift within the FRII source population. The source distribution in
the P-D plane predicted by our model is compared to that of the
observed LRL sample. We find that our model predicts an unphysically
strong evolution of the gas density in the source environment with
redshift, if we assume that all jet and environment parameters are
independent of each other. The best fit of the model to the data in
the P-D plane is achieved by assuming that the life time of radio
sources or the shape of the density distribution of their environments
depends on the power of their jets.

Using this approach we find evidence that the giant sources in the LRL
sample with linear sizes greater than 1.5 Mpc, DA240, 3C236 and
3C326, constitute a class of objects intrinsically different from
the rest of the sample. They have to be extremely old and/or are
located in extremely underdense environments. 

The luminosity function of FRII sources derived from the models is in
good agreement with the results of Dunlop \& Peacock
(1990)\nocite{dp90}. We find evidence for a flattening of the
luminosity function beyond $z=1.5$. The evolution of the luminosity
function is dominated by density evolution but pure density evolution
is ruled out because of the inverse Compton scattering losses of the
CMBR off the relativistic electrons in the cocoons of radio
sources. This result only applies to FRII sources with the highest
radio luminosities at any given cosmological epoch which comprise the
LRL sample. For lower luminosities the cosmological evolution of the
FRII source population is probably different and the models presented
here have to be modified to allow for such a behaviour of the radio
luminosity function. Fainter complete samples of radio sources are
also needed to better constrain the models at lower luminosities. The
cosmological linear size evolution predicted by the models is
consistent with the weak evolution derived by Neeser {\em et al.}
(1995)\nocite{ner95} from observations.

The best-fitting model predicting a correlation of the jet power with
the source life time, model F, explains the decline of the radio
luminosity of the most luminous radio galaxies with cosmic epoch in
terms of a decrease in the efficiency with which the jets are
fuelled. The alternative model G requires cosmological evolution not
only of the central density but also of the core radius of the density
distribution of the material surrounding radio galaxies. This is
consistent with the observed change in the environments of radio
galaxies with cosmic epoch. In this model, the decline of the radio
luminosity of the most luminous sources is caused by the decreasing
mass of the central black holes powering the jets.

\section*{Acknowledgements}

We thank S. Rawlings, M. Lacy and the anonymous referee for helpful
comments on the manuscript.

\bibliography{../../crk}

\begin{thebibliography}{Kaiser, Dennett-Thorpe \& Alexander<1997>}

\bibitem[Alexander \& Leahy<1987>]{al87}
Alexander~P., Leahy~J.~P., 1987, MNRAS, 225, 1

\bibitem[Baldwin {\rm et~al.}<1985>]{bbhjwww85}
Baldwin~J.~E., Boysen~R.~C., Hales~S. E.~G., Jennings~J.~E., Waggett~P.~C.,
  Warner~P.~J., Wilson~D. M.~A., 1985, MNRAS, 217, 717

\bibitem[Baldwin<1982>]{jb82}
Baldwin~J.~E., 1982, in Heeschen~D.~S., Wade~C.~M., eds, Extragalactic radio
  sources.
\newblock Reidel, p.~21

\bibitem[Barthel<1989>]{pb89}
Barthel~P.~D., 1989, ApJ, 336, 606

\bibitem[Best, Longair \& R{\" o}ttgering<1998>]{blr98}
Best~P.~N., Longair~M.~S., R{\" o}ttgering~H. J.~A., 1998, MNRAS, 295, 549

\bibitem[Canizares, Fabbiano \& Trinchieri<1987>]{cft87}
Canizares~C.~R., Fabbiano~G., Trinchieri~G., 1987, ApJ, 312, 503

\bibitem[Cohen \& Ryzhov<1997>]{cr97}
Cohen~J.~G., Ryzhov~A., 1997, ApJ, 486, 230

\bibitem[Dunlop \& Peacock<1990>]{dp90}
Dunlop~J.~S., Peacock~J.~A., 1990, MNRAS, 247, 19

\bibitem[Eales<1985>]{se85}
Eales~S.~A., 1985, MNRAS, 217, 179

\bibitem[Efstathiou \& Rees<1988>]{er88}
Efstathiou~G., Rees~M.~J., 1988, MNRAS, 230, 5p

\bibitem[Ellingson, Green \& Yee<1991>]{egy91}
Ellingson~E., Green~R.~F., Yee~H. K.~C., 1991, ApJ, 378, 476

\bibitem[Heavens \& \protect{O'C}. Drury<1988>]{hd88}
Heavens~A.~F., \protect{O'C}. Drury~L., 1988, MNRAS, 235, 997

\bibitem[Hill \& Lilly<1991>]{hl91}
Hill~G.~J., Lilly~S.~J., 1991, ApJ, 367, 1

\bibitem[Jones \& Forman<1984>]{jf84}
Jones~C., Forman~W., 1984, ApJ, 276, 38

\bibitem[Kaiser \& Alexander<1997>]{ka96b}
Kaiser~C.~R., Alexander~P., 1997, MNRAS, 286, 215

\bibitem[Kaiser, Dennett-Thorpe \& Alexander<1997>]{kda97a}
Kaiser~C.~R., Dennett-Thorpe~J., Alexander~P., 1997, MNRAS, 292, 723

\bibitem[Kapahi<1989>]{vk89}
Kapahi~V.~K., 1989, AJ, 97, 1

\bibitem[King<1972>]{ik72}
King~I.~R., 1972, ApJ, 174, L123

\bibitem[Kormendy \& Richstone<1995>]{kr95}
Kormendy~J., Richstone~D., 1995, ARA\&A, 33, 581

\bibitem[Laing, Riley \& Longair<1983>]{lrl83}
Laing~R.~A., Riley~J.~M., Longair~M.~S., 1983, MNRAS, 204, 151

\bibitem[Leahy \& Williams<1984>]{lw84}
Leahy~J.~P., Williams~A.~G., 1984, MNRAS, 210, 929

\bibitem[Macklin<1982>]{jm82}
Macklin~J.~T., 1982, MNRAS, 199, 1119

\bibitem[Masson<1980>]{cm80}
Masson~C.~R., 1980, ApJ, 242, 8

\bibitem[Neeser {\rm et~al.}<1995>]{ner95}
Neeser~M.~J., Eales~S.~A., Duncan-Green~J., Leahy~J.~P., Rawlings~S., 1995,
  ApJ, 451, 76

\bibitem[Nulsen \& B{\" o}hringer<1995>]{nb95}
Nulsen~P. E.~J., B{\" o}hringer~H., 1995, MNRAS, 274, 1093

\bibitem[Oort {\rm et~al.}<1987>]{okw87}
Oort~M. J.~A., Katgert~P., Steeman~F. W.~M., Windhorst~R.~A., 1987, A\&A, 179,
  41

\bibitem[Peacock<1983>]{jp83}
Peacock~J.~A., 1983, MNRAS, 202, 615

\bibitem[Peacock<1985>]{jp85}
Peacock~J.~A., 1985, MNRAS, 217, 601

\bibitem[Rawlings \& Saunders<1991>]{rs91}
Rawlings~S., Saunders~R., 1991, Nat., 349, 138

\bibitem[Riley<1989>]{jr89}
Riley~J.~M., 1989, MNRAS, 238, 1055

\bibitem[Shklovskii<1963>]{is63}
Shklovskii~I.~S., 1963, SvA, 6, 465

\bibitem[Singal<1988>]{as88}
Singal~A.~K., 1988, MNRAS, 233, 87

\bibitem[Subrahmanyan \& Swarup<1990>]{ss90}
Subrahmanyan~K., Swarup~G., 1990, MNRAS, 247, 237

\end{thebibliography}
\bibliographystyle{../../mnras}

\end{document}